\documentclass[12pt]{article}
\usepackage{graphicx}
\usepackage{cite}
\usepackage{amsmath}
\usepackage{color}
\usepackage{hyperref}
\usepackage{authblk}
\usepackage{subfigure}
\usepackage{comment}
\usepackage[titletoc,title]{appendix}
\usepackage{booktabs} 
\usepackage[labelfont=bf]{caption} 

\definecolor{darkblue}{rgb}{0,0,.7}
\definecolor{darkred}{rgb}{0.7,0,0}

\hypersetup{colorlinks=true, linkcolor=darkblue,citecolor=darkblue,urlcolor=darkblue}

\numberwithin{equation}{section}

\title{Radiative corrections to neutron and nuclear $\beta$-decays: a serious kinematics problem in the 
literature}

\author[1]{Ferenc  Gl\"uck\thanks{ferenc.glueck@kit.edu}\thanks{I dedicate this paper to the memory of A. Sirlin and 
A. N. Ivanov}}
\affil[1]{Karlsruhe Institute of Technology, IAP, 76021 Karlsruhe, POB 3640, Germany}

\date{}

\begin{document}

\maketitle

\begin{abstract}

We report a serious kinematics problem in the bremsstrahlung photon part of the 
order-$\alpha$ outer (model independent) radiative correction calculations for those neutron (and nuclear beta) decay observables
(like electron-neutrino correlation parameter measurement) where the proton (recoil particle) is detected.
The so-called neutrino-type radiative correction calculations, which fix the neutrino direction in the
bremsstrahlung photon integrals, 
use 3-body decay kinematics to connect the unobserved neutrino direction with the
observed electron and proton (recoil particle) momenta. 
But the presence of the bremsstrahlung photon changes the kinematics from 3-body to 4-body one,
and the accurate information about the recoil particle momentum is lost
due to the integration with respect to the photon momentum.
Therefore the application of the abovementioned 3-body decay kinematics connection for the radiative correction calculations,
rather prevalent in the literature, is not acceptable.
We show that the correct, so-called recoil-type  radiative correction calculations, which fix the proton (recoil particle)
momentum instead of the neutrino direction and use
rather involved analytical, semianalytical or
Monte Carlo bremsstrahlung integration methods, result usually in much larger corrections than the incorrect
neutrino-type analytical methods.
\end{abstract}

\medskip

\newpage

\tableofcontents

\section{Introduction}
 \label{SectionIntroduction}

High precision experiments and analyses of beta and semileptonic decays provide important possibilities
to test the Standard Model (SM) of particle physics, e.g. to test the unitarity of the CKM matrix or the presence of non-standard
right-handed, scalar or tensor couplings.
In order to test the small deviations from the SM, the analyses of these
experiments require to use precise and reliable radiative corrections to the theoretical distributions, rates and asymmetries.
Radiative corrections have been very important for the development and  verification of the Standard Model
of elementary particles, especially for the V-A and quark mixing part of the SM \cite{SirlinFerroglia2013}.
Most likely, the confirmation or falsification of the SM will depend also in the future on precise and reliable
radiative correction calculations (as well as on precise and reliable measurements).

The QED part of the electroweak radiative corrections is connected with
transition amplitudes  containing virtual and bremsstrahlung photons in the Feynman graphs.
The off-shell virtual
photons are created and absorbed by the charged participants of the beta decay, therefore
the transitions with virtual photons have the same
3-body kinematics as the zeroth-order transitions.
The bremsstrahlung photon is emitted by one of the charged particles (e.g. the electron) and,
contrary to the virtual photon, it leaves the beta decay region, and therefore it results in 4-body
decay kinematics.

In order to measure the electron-neutrino correlation parameter $a$ or the neutrino asymmetry parameter
$B$ in beta decays, the neutrino momentum has to be determined. Since the neutrino is usually not detected,
the neutrino momentum can be  reconstructed by measuring the electron and recoil particle (proton) momenta.
Without the bremsstrahlung photon, the beta decay has 3-body kinematics, and in this case 
the electron-neutrino angle or the neutrino direction can be determined by measuring the electron and
recoil particle momenta. In the presence of bremsstrahlung photon, one has 4-body decay kinematics
(See Fig. \ref{FigKinematics} in Sec. \ref{SectionKinematics}), and 
the neutrino momentum determination is not possible (assuming that neither the photon nor the neutrino
are detected); 
one can determine only the 
sum of the neutrino and bremsstrahlung photon momenta. It is important that the radiative correction
calculations should  use only the observable electron and recoil particle momenta, but not the
unobserved neutrino momentum, as fixed quantities; the unobserved neutrino and photon momenta should be
integrated over the phase space, according to the kinematical constraints determined by the experimental details.
E.g. one could define the neutrino + bremsstrahlung
photon momentum sum as a so-called pseudo-neutrino momentum: in the radiative correction
calculations for the electron-neutrino correlation or neutrino asymmetry parameters, one 
should use this pseudo-neutrino momentum, instead of the real neutrino momentum, as fixed quantity.

Radiative correction calculations to muon, nuclear (neutron) and pion beta decay were started already in the middle of the
fifties, i.e. much earlier than the SM was established; in fact,  these results were important for the development of the SM
(e.g. for the introduction of the Cabibbo angle). As an example, a simple analytical formula for the radiative correction to the electron energy
spectrum of beta decay was presented by Kinoshita and Sirlin in Ref. \cite{KinoshitaSirlin1959} (Eq. 4.1).
In the late sixties and early seventies, Ginsberg and others published radiative correction results for various quantities 
(like Dalitz distributions) in K meson decays \cite{Ginsberg1966,Ginsberg1967,Ginsberg1968,Ginsberg1970,Becherrawy1970,KamalWong1971}. Sirlin introduced
 in 1967 \cite{Sirlin1967} (and later established within the SM framework \cite{SirlinFerroglia2013})
the concept of subdividing the order-$\alpha$ radiative correction into inner (model dependent)
and outer (model independent, MI) parts: the inner part is connected with the high-energy virtual photons and electroweak 
$W$ and $Z$ bosons, while the outer part is the sum of the low-energy virtual photon and bremsstrahlung photon corrections.
The outer correction, which has only small strong interaction dependence in low energy beta decays, can be rather reliably computed, 
it changes the spectrum shapes and asymmetries, and is sensitive to experimental details \cite{Gluck2003}.
On the other hand, the inner correction  does not 
change the energy spectrum shapes 
(with good approximation in neutron and low energy nuclear decays), and can be absorbed into the weak vector and axialvector couplings.
The main subject of our paper is the bremsstrahlung and thus the outer radiative corrections, and we do not deal at all with the
inner corrections;
see Refs. \cite{Seng2019,Czarnecki2019,Hayen2021} for recent publications about the inner radiative corrections
to neutron and nuclear beta decays.

Sirlin presented in his seminal paper of Ref. \cite{Sirlin1967}   the 
outer (MI) radiative correction to the electron energy spectrum of
nuclear beta decays as a simple and universal analytical function: $g(E_2)$ of Eq. 20 in Ref. \cite{Sirlin1967},
where $E_2$ denotes the electron energy;
 see also Eq. 2.7 in Ref. \cite{Gluck1993}, Eq. 42 in Ref. \cite{SirlinFerroglia2013}, and Eq. \ref{gE2} in appendix 
\ref{SectionRadCorrFormulas}.
 The outer radiative correction to the electron
asymmetry in polarized neutron decay \cite{Markisch2019} was presented by Shann in 1971 
(using the function $g(E_2)$ of Sirlin and the function $h(E_2)$\footnote{Note that this has nothing to do with the function $h$
of Eq. 49 in Ref. \cite{SirlinFerroglia2013} and Eq. 11 in Ref. \cite{Sirlin2011}, which is a simple analytical result for the radiative correction of the 
neutrino energy spectrum in beta decays.}
of Eq. 10 in Ref. \cite{Shann1971}; see Eqs. \ref{gE2}, \ref{fn}).
It is important to mention that in both observables, i.e. the electron energy spectrum of Sirlin and the electron asymmetry of Shann,
the recoil particle (proton in the case of neutron decay) is not detected. For the radiative correction calculations 
this corresponds
to complete integration with respect to the recoil particle momentum, and this enables the integration over the
bremsstrahlung photon phase space by relatively simple analytical methods. If one fixes the recoil particle energy (momentum), then
the integration limits are more complicated and the analytical integrations are much more difficult. Note that, due to the large
electron-proton (or electron-nucleus) mass difference, the radiative correction calculations to the electron energy spectrum
 and to the recoil 
particle (proton) energy spectrum are completely different (both from the kinematics and from the dynamics point of view), 
and this is why the relatively simple analytical integration for the electron spectrum
radiative correction  is 
possible, while for the recoil particle spectrum this is not possible  (see Sec. \ref{SectionBRIntegral} for more details). 

Several authors observed in the seventies 
\cite{YokooMorita1976,FujikawaIgarashi1976,GarciaMaya1978,Garcia1978}
that the analytical integrations of the bremsstrahlung corrections remain simple even if one fixes at the integration, 
in addition to the electron energy
and direction, also the neutrino direction (but not the neutrino energy). The authors claimed then that their analytical
formulas provide radiative correction results for electron-neutrino correlation and neutrino asymmetry
quantities. In fact, these extended radiative correction formulas can be expressed by using the abovementioned analytical
functions $g(E_2)$ introduced by Sirlin and $h(E_2)$ introduced by Shann.
In the beginning of the eighties, these analytical methods were extended to the radiative correction
calculations of hyperon semileptonic decays
\cite{Garcia1981,Garcia1982}.

Nevertheless, in 1984 K\'alm\'an T\'oth pointed out in Ref. \cite{Toth1984} that the analytical radiative correction 
results of Refs. \cite{YokooMorita1976,FujikawaIgarashi1976,GarciaMaya1978,Garcia1978,Garcia1981,Garcia1982}
 are not appropriate for the precision  electron-neutrino
correlation analyses of semileptonic decays. Namely, these calculations assume a connection between the neutrino and
recoil particle momentum by using 3-body kinematics, and due to the 
presence of bremsstrahlung photons this is not a good approximation.
In fact, the radiative correction results of Refs. 
\cite{YokooMorita1976,FujikawaIgarashi1976,GarciaMaya1978,Garcia1978,Garcia1981,Garcia1982}
can be applied only for the analyses of hypothetical experiments with neutrino detection;
for most of the beta decay experiments this is of course not the case.

The radiative corrections to the electron and recoil energy Dalitz distributions for
neutron and several hyperon semileptonic decays were computed in Refs.
 \cite{ChristianKuhnelt1978,TothMargaritisSzego1984,TothSzegoMargaritis1986}
by numerical integrations over the bremsstrahlung photon phase space.
Afterwards rather elaborate analytical methods  of the 3-dimensional bremsstrahlung integrations
were developed in Refs. \cite{Nagy,Gluck1986-89}, and
the radiative correction results obtained with these methods were
published in Refs. \cite{TothGluck1989,GluckToth1990,GluckToth1992,Gluck1993}.
The above described bremsstrahlung kinematics issue and the inadequacy of
the radiative correction results of Refs. 
 \cite{YokooMorita1976,FujikawaIgarashi1976,GarciaMaya1978,Garcia1978,Garcia1981,Garcia1982}
for the beta decay experimental analyses
were explained in many of our publications \cite{TothGluck1989,GluckToth1990,GluckToth1992,Gluck1993,
GluckJoo1994, Gluck1997,Gluck1998a,Gluck1998b,Gluck2003}
and talks \cite{Gluck2007,Gluck2014}.

The authors of Refs. \cite{Ginsberg1966,Ginsberg1967,Ginsberg1968,Ginsberg1970,Becherrawy1970,KamalWong1971}
 used already in the sixties and early seventies the correct brems\-strahlung integration methods
for the Dalitz-plot radiative correction calculations of K-meson decays. The word ``correct'' here means: by using explicitly 
the recoil particle (and the electron) as the observable, instead of fixing the direction unit vector of the  undetected neutrino.
We introduce now the following terminology: we call  recoil-type radiative correction\footnote{This is completely different from
recoil-order corrections, which are  small (order of recoil particle kinetic energy divided by beta decay energy)
improvements of the zeroth-order (or the radiative correction) calculations.}
calculations where, in addition to the electron momentum, the recoil particle
momentum is fixed in the bremsstrahlung integrals \cite{Ginsberg1966,Ginsberg1967,Ginsberg1968,Ginsberg1970,Becherrawy1970,KamalWong1971,ChristianKuhnelt1978,TothMargaritisSzego1984,
TothSzegoMargaritis1986,TothGluck1989,GluckToth1990,GluckToth1992,Gluck1993,
GluckJoo1994, Gluck1997,Gluck1998a,Gluck1998b},
and neutrino-type radiative correction calculations where the neutrino
direction is fixed \cite{YokooMorita1976,FujikawaIgarashi1976,GarciaMaya1978,Garcia1978,Garcia1981,Garcia1982}.
After the critical publications of T\'oth et al, from the end of the eighties the radiative
correction calculations for hyperon semileptonic decays 
(Refs. \cite{Tun1989,Tun1991,Martinez1993,Juarez1993,Martinez2000,FloresMendieta2006}) and for K meson semileptonic decays
(Refs. \cite{Cirigliano2002,Bytev2003,Cirigliano2004,Andre2005,Andre2007,JuarezLeon2011,Neri2015,Seng2021a,Seng2021b,Seng2022})
have  been using the correct recoil-type integration methods (this list of radiative correction publications
is far from complete).
Unfortunately, the situation is different in the case of neutron decay: in the past two decades, most of the 
radiative correction calculation publications 
\cite{Ando2004,Gudkov2005,Gudkov2006,IvanovPitschmann2013,IvanovPitschmannarXiv2018,
IvanovHollwieser2013,IvanovHollwieser2019,Ivanov2021}
used the incorrect neutrino-type integration method. One has to emphasize here that these radiative correction results
are inappropriate only for those neutron decay experiments where the proton is detected (like measurements of the
electron-neutrino correlation or neutrino asymmetry parameter).
The neutrino-type radiative correction results are suitable for observables (like electron energy
spectrum, lifetime or electron asymmetry) where the proton is not detected
(in the sense that the proton detection is not necessary for the experimental determination of the observable); 
and in that case the recoil-type and neutrino-type
radiative correction results are practically equal (the very small differences in their results are due to different recoil-order
terms in the radiative correction quantities).

An important quality requirement of the outer radiative correction calculations and their applications
in the experimental analyses is consistency: the radiative correction calculation conditions (details)  have to be consistent
with the experimental conditions. Put in other words, the radiative correction calculations 
have to take into account correctly the experimental details,
 i. e. which observables of the beta decay particles are measured,
in which  intervals etc?
For instance, the radiative corrections to the electron and proton energy Dalitz
distributions can be applied to the neutron decay experiments (e.g. Nab \cite{Nab2017})
where only these two quantitities are measured, without any
experimental constraints about the angle between these particles. 
One has to emphasize that
the outer radiative corrections are not unique:
different measurement methods require different radiative correction calculations.
For instance, 
in order to determine the electron-neutrino correlation parameter $a$ in neutron decay,
there are several different measurement methods:
electron and proton energy Dalitz distribution \cite{Nab2017}; 
proton energy spectrum without electron detection \cite{aSPECT2020}; measurement of the 
electron and proton momenta (i.e. both energy and direction) \cite{aCORN2021}
etc. The analyses of these various experiments require completely different 
radiative correction results, so we cannot speak about \textbf{the} radiative correction calculation to
the electron-neutrino correlation parameter determination.

In the case of the neutrino-type radiative correction calculation, one performs the integration over the bremsstrahlung photon
phase space by fixing the electron energy and the electron and neutrino directions
(the neutrino energy is strongly correlated with the electron and photon energy).
The first problem here is the neutrino direction fixing, since the usual beta decay experiments do not detect the
neutrinos.
In order to circumvent this problem,  the authors of these calculations
make a connection between the electron and recoil particle parameters and the neutrino direction 
by using zeroth-order 3-body kinematics.
This is the second  problem of the neutrino-type radiative corrections: during the integration over the bremsstrahlung photon
phase space, the recoil particle momentum changes together with the photon momentum (due to momentum conservation),
and therefore the information about the recoil particle momentum is lost after this integration.
In other words, the neutrino-type radiative corrections have a serious consistency problem: 
the neutrino direction is used as fixed quantity in the calculations, while it is not used in the experimental analyses at all
(since the neutrino is not detected);
 and the observable recoil particle momentum or energy is used as fixed quantity in the experimental analyses, 
but in the radiative calculations it is integrated together with the bremsstrahlung photon, and so it 
is not used as fixed quantity (or it is used in a different way as in the experiment).

We would like to emphasize that, in spite of our serious critics about the handling of the bremsstrahlung integrations,
we appreciate very much the authors of the neutrino-type radiative correction publications
 \cite{YokooMorita1976,FujikawaIgarashi1976,GarciaMaya1978,Garcia1978,Garcia1981,Garcia1982,
    Ando2004,Gudkov2005,Gudkov2006,IvanovPitschmann2013,
IvanovPitschmannarXiv2018,IvanovHollwieser2013,IvanovHollwieser2019,Ivanov2021},
because these papers contain a lot of  precious and useful scientific merits;
e. g. we have made many comparisons of our results with these publications, and in most cases
we have found good agreement.

The main goal of our present paper is to provide a new and somewhat more
detailed explanation about the above described  problem of radiative correction calculations.
In Sec. \ref{SectionKinematics} we explicate the 3-body beta decay kinematics without photon and the
4-body kinematics with bremsstrahlung photon, exhibit the difference of the
 electron and recoil particle energy Dalitz plots without and with bremsstrahlung photons, and
outline some properties of the bremsstrahlung photon parameters.
Sec. \ref{SectionBRIntegral} deals with the technically most difficult part of the radiative corrections: the 
bremsstrahlung integral. We show different recoil-type calculation methods of this integral, and we explain why 
is the neutrino-type calculation method inadequate for the experimental analyses.
In Sec. \ref{SectionComparison} we compare the recoil-type and neutrino-type radiative correction results of various
distributions: electron and recoil particle energy Dalitz plot, recoil particle energy spectrum and electron-neutrino correlation Dalitz 
distribution. The plots presented here reveal that these two correction results are completely different
(for the observables with detected recoil particle), and usually the
recoil-type corrections are much larger than the neutrino-type corrections.

\section{3-body and 4-body decay kinematics}
 \label{SectionKinematics}

\begin{figure}[!htbp]
    \centering
    \includegraphics[width=0.75\textwidth]{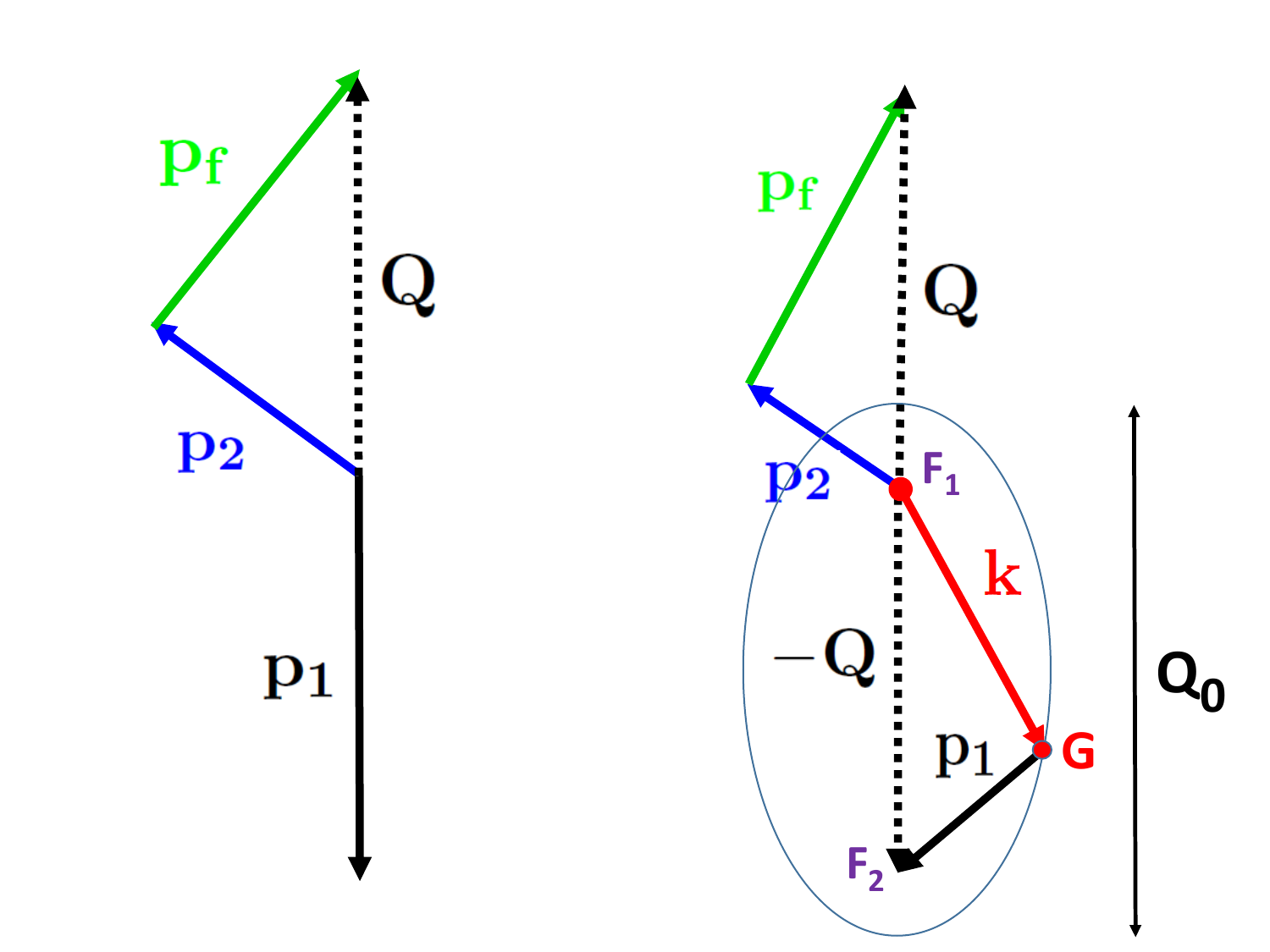}
    \caption{3-body decay kinematics without bremsstrahlung photon (left) and 4-body kinematics 
                with bremsstrahlung photon (right), with electron ( ${\bf p_2}$), recoil particle ( ${\bf p_f}$),
               neutrino ( ${\bf p_1}$) and photon ( ${\bf k}$) momentum vectors.}
    \label{FigKinematics}
\end{figure}

Figure \ref{FigKinematics} shows the beta decay kinematics without and with bremsstrahlung photon 
(left and right, respectively). We can see there the momentum vectors of the electron ${\bf p_2}$,
the recoil (final hadron or nucleus) particle ${\bf p_f}$, the neutrino ${\bf p_1}$ and the photon ${\bf k}$, and additionally
the  vector ${\bf Q}={\bf p_2}+{\bf p_f}$.
In neutron or nuclear beta decay (or semileptonic beta decay) experiments, usually neither the neutrino nor the bremsstrahlung photon
are detected, therefore one has experimental information only about the $({\bf p_2},{\bf p_f},{\bf Q})$
decay triangle in the upper parts of the figure.
Without bremsstrahlung photon we have 3-body kinematics, with  momentum conservation in the
decaying particle CMS: ${\bf p_2}+{\bf p_f}+{\bf p_1}=0$.
From the 9 parameters of the 3 
momentum vectors, due to the 4 energy and momentum conservation equations and the 3 Euler angles
for the unpolarized case, the kinematics decay triangle $({\bf p_2},{\bf p_f},{\bf Q})$
is defined by 9-4-3=2 parameters, e.g. by the electron total (mass+kinetic) energy $E_2$ and the
recoil particle kinetic energy $T=E_f-m_f$ (we use the usual paricle physics $c_{\rm light}=1$ convention in our paper,
therefore the mass has also eV unit like the energy). 
The neutrino is usually
not measured, but using the electron and recoil particle detection parameters and  the 3-body kinematics,
the neutrino parameters (e.g. its angle to the electron) can be determined.
We mention that in some cases one has a twofold ambiguity for the $({\bf p_2},{\bf p_f},{\bf Q})$ decay
triangle determination from 2 measured parameters; e.g. if the electron energy and the angle between the electron and
the recoil particle are measured, and the electron energy is large, there are 2 solutions for the recoil particle energy, and thus for the
$({\bf p_2},{\bf p_f},{\bf Q})$ decay triangle (see Sec. 5 in Ref. \cite{Gluck1993}; one can also see on Fig. \ref{FigDalitzPlot3} 
that for $E_2>E_{2h}$ and for $180^\circ$ between ${\bf p_2}$ and ${\bf p_f}$ there are two solutions for $T$ with fixed $E_2$).

In the case of the presence of bremsstrahlung photons we have 4-body kinematics with 12 momentum
parameters, but the same number of conservation equations and Euler angles as above, therefore
12-4-3=5 parameters determine the unpolarized decay kinematics; e.g. these could be the electron
and recoil particle energy, the third side of the decay triangle $Q=|{\bf Q}|$, the photon momentum magnitude
$K=|{\bf k}|$, and the azimuthal angle $\phi_k$ of the photon momentum vector ${\bf k}$ around
 the vector ${\bf Q}$ \cite{TothSzegoMargaritis1986,GluckToth1990}. 
The kinematical limits of these 4-body decay parameters can be determined by the energy and momentum
conservation equations:
\begin{equation} \label{ConservationEquations}
Q_0=\Delta-E_2-T=E_1+K, \quad {\bf p_2}+{\bf p_f}={\bf Q}=-{\bf p_1}-{\bf k},
\end{equation}
where $\Delta=m_i-m_f$ denotes the mass difference of the decaying (initial) and daughter (final) nuclei, and
$E_1=|{\bf p_1}|$ is the neutrino energy. For a fixed $({\bf p_2},{\bf p_f},{\bf Q})$ decay triangle,
both the $E_1+K$ scalar sum and the ${\bf p_1}+{\bf k}$ vector sum have to be fixed. Therefore,
the endpoint $G$ of the photon momentum vector ${\bf k}$ in Fig. \ref{FigKinematics}
(same as the starting point of vector ${\bf p_1}$) can move on the surface of a rotational ellipsoid
with major axis $Q_0$ and 
focal points $F_1$ and $F_2$ whose distance is $Q$ \cite{Gluck1986-89}. 
In this case,  the electron and recoil particle energy and the angle between their 
directions are needed for the complete determination of the $({\bf p_2},{\bf p_f},{\bf Q})$ decay triangle, 
and even then, one cannot 
ascertain the neutrino parameters (energy and direction); at the most the  ${\bf p_1}+{\bf k}$ momentum vector sum
and the $E_1+K$ energy sum (momentum and energy of the pseudo-neutrino, defined in Sec. \ref{SectionIntroduction})
can be determined.

\begin{figure}[htbp]
    \centering
    \subfigure[neutron decay]{\includegraphics[width=0.47\textwidth]
                  {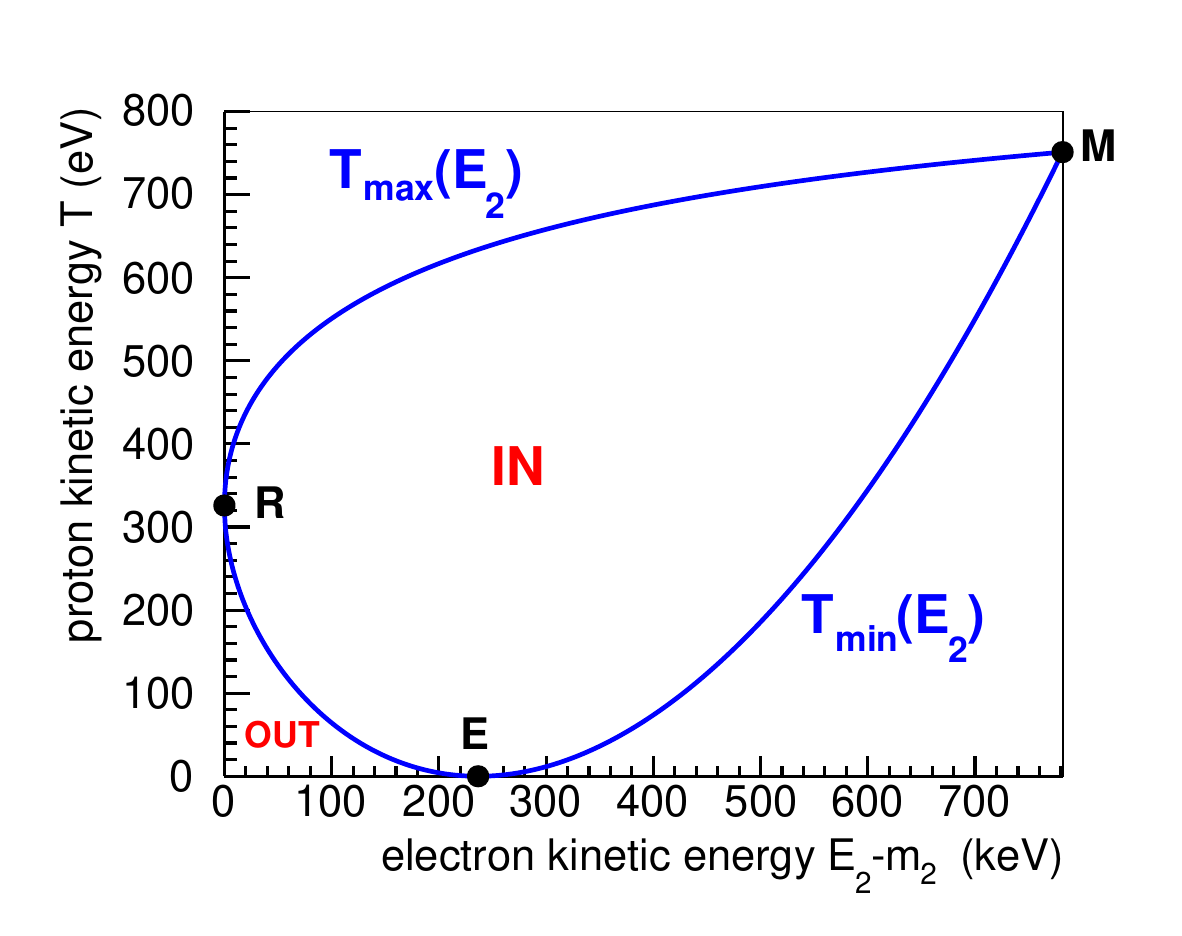}  \label{FigDalitzPlot1}}\quad
    \subfigure[10 MeV nuclear decay]{\includegraphics[width=0.47\textwidth]
                  {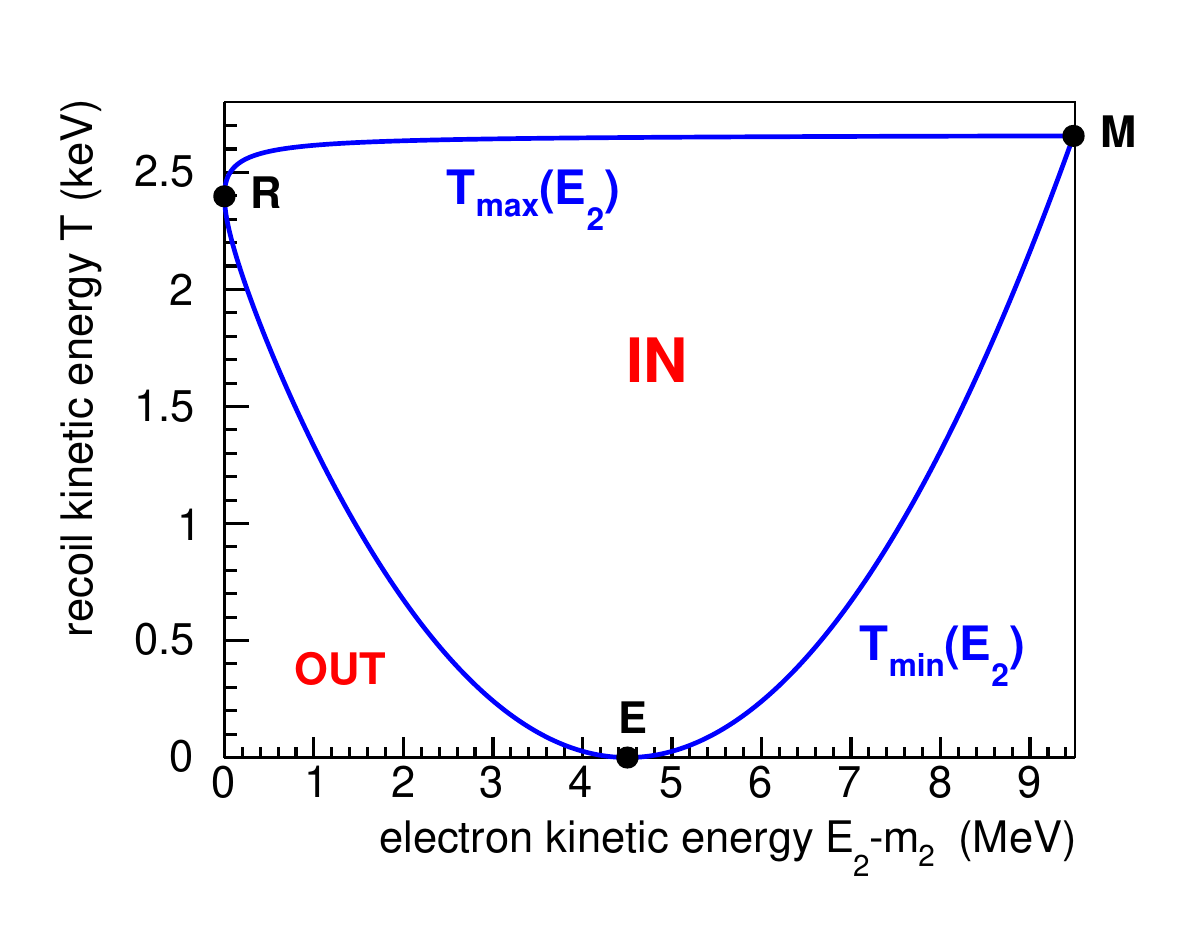}    \label{FigDalitzPlot2}}         
    \caption{$(E_2-m_2,T)$ Dalitz plots for neutron decay (left) and a nuclear beta${}^-$ decay with $\Delta=10$ MeV
                  and $m_f=20\; m_p$ (proton mass).}
    \label{FigDalitzPlots}
\end{figure}

Fig. \ref{FigDalitzPlots} shows the electron and recoil particle energy $(E_2,T)$ Dalitz plots for the neutron decay
and for a hypothetical nuclear beta decay with $\Delta=10$ MeV and $m_f=20\; m_p$ (proton mass).
The Dalitz boundary formulas $T_{min/max}(E_2)$ and the parameters of the special points
$R$, $E$ and $M$ can be found in App.  \ref{SectionDalitzPlotFormulas}
(e. g. the point E is at $T=0$ and $E_2=E_{2h}$ defined by Eq. \ref{ERpoints}).
In the case of 3-body decay kinematics, the allowed Dalitz region is defined by the electron energy limits $m_2<E_2<E_{2m}$
and by the recoil particle energy limits $T_{min}(E_2)<T<T_{max}(E_2)$;  we call this region IN.
Fig. \ref{FigDalitzPlot3} in  App.  \ref{SectionDalitzPlotFormulas} shows that, with 3-body decay kinematics, 
the ${\bf p_2}$, ${\bf p_f}$  and ${\bf p_1}$ momentum vectors are parallel or antiparallel to each other on the
Dalitz-plot boundaries $T=T_{min}(E_2)$ and $T=T_{max}(E_2)$.
In the case of 4-body decay kinematics (in the presence of bremsstrahlung photons),
the region OUT, defined by $0<T<T_{min}(E_2)$ for $E_2<E_{2h}$
(the region between the axes and the R-E curve in the left-bottom corner of Fig. \ref{FigDalitzPlots}) is also allowed,
in addition to the region IN (see Refs. 
\cite{TothSzegoMargaritis1986,Gluck1986-89,GluckToth1990,Tun1989,Cirigliano2004,JuarezLeon2011}).

Using Fig. \ref{FigDalitzPlot3} in App.  \ref{SectionDalitzPlotFormulas},
one can easily understand why do the boundaries R-M and E-M not change in the transition from 3-body to 4-body decay
kinematics. E.g. in the case of curve R-M, in order to extend upwards this boundary in the case of 4-body kinematics, i.e. to increase the $P_f=|{\bf p_f}|$
recoil particle momentum, the bremsstrahlung momentum vector ${\bf k}$ has to be opposite to ${\bf p_f}$, i. e.
parallel to the neutrino momentum ${\bf p_1}$. But, in our approximation, both the neutrino and photon are massless, 
so the momentum vector ${\bf p_1}+{\bf k}$ has the same kinematical properties in 4-body kinematics as ${\bf p_1}$
in 3-body kinematics,
and therefore $P_f$ has the same maximum values in both cases. Similar arguments hold for the curve E-M.
In the case of the boundary R-E the situation is different: the photon momentum ${\bf k}$ opposite to the neutrino momentum
${\bf p_1}$ can replace the momenta ${\bf p_2}$ and ${\bf p_f}$, and so, in the presence of bremsstrahlung photon, 
the electron and recoil particle kinetic energies can be
reduced relative to the values on the boundary curve R-E.

\begin{figure}[htbp]
    \centering
    \subfigure[IN region]{\includegraphics[width=0.47\textwidth]
                  {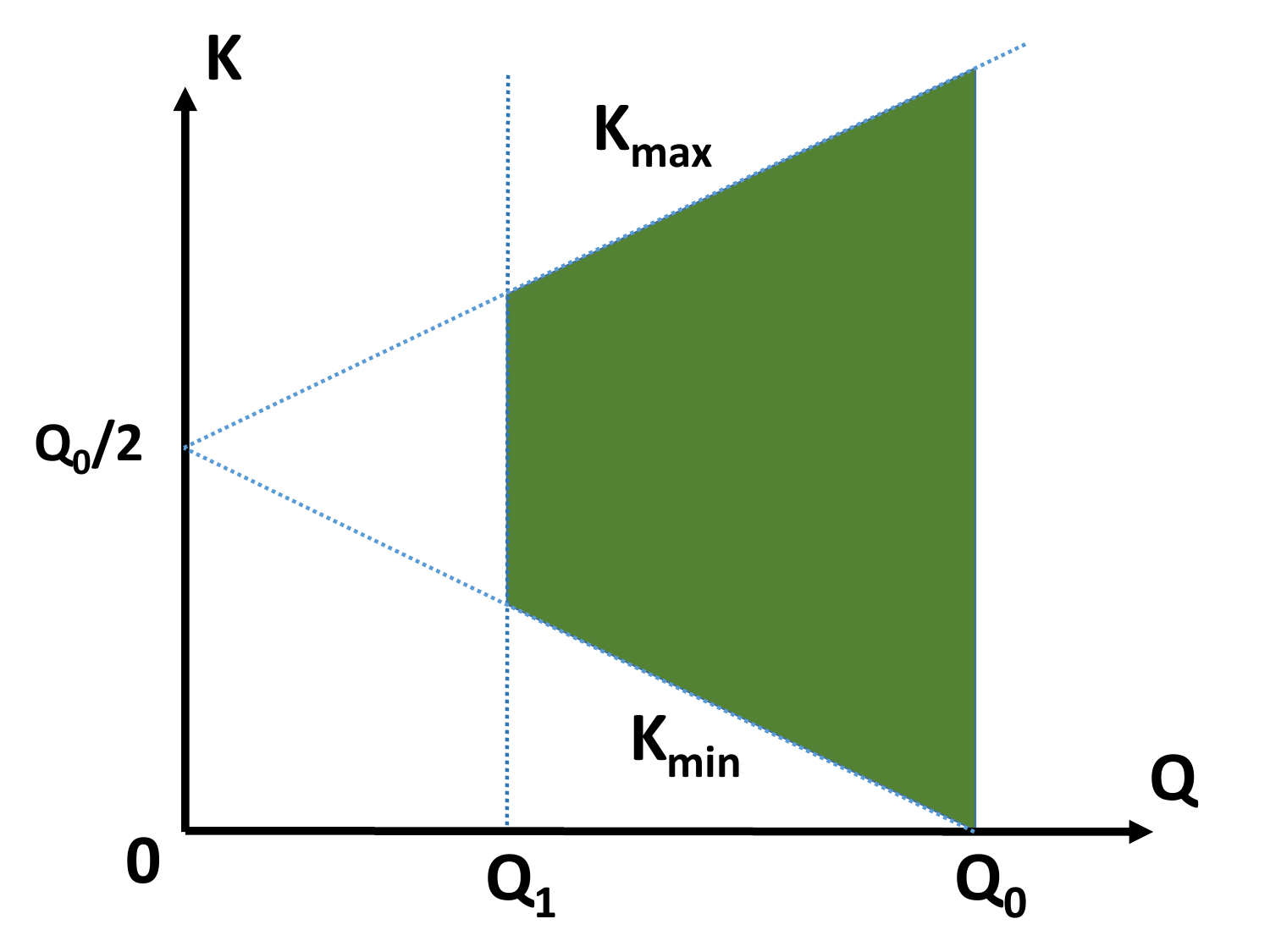}  \label{FigQKplotA}}\quad
    \subfigure[OUT region]{\includegraphics[width=0.47\textwidth]
                  {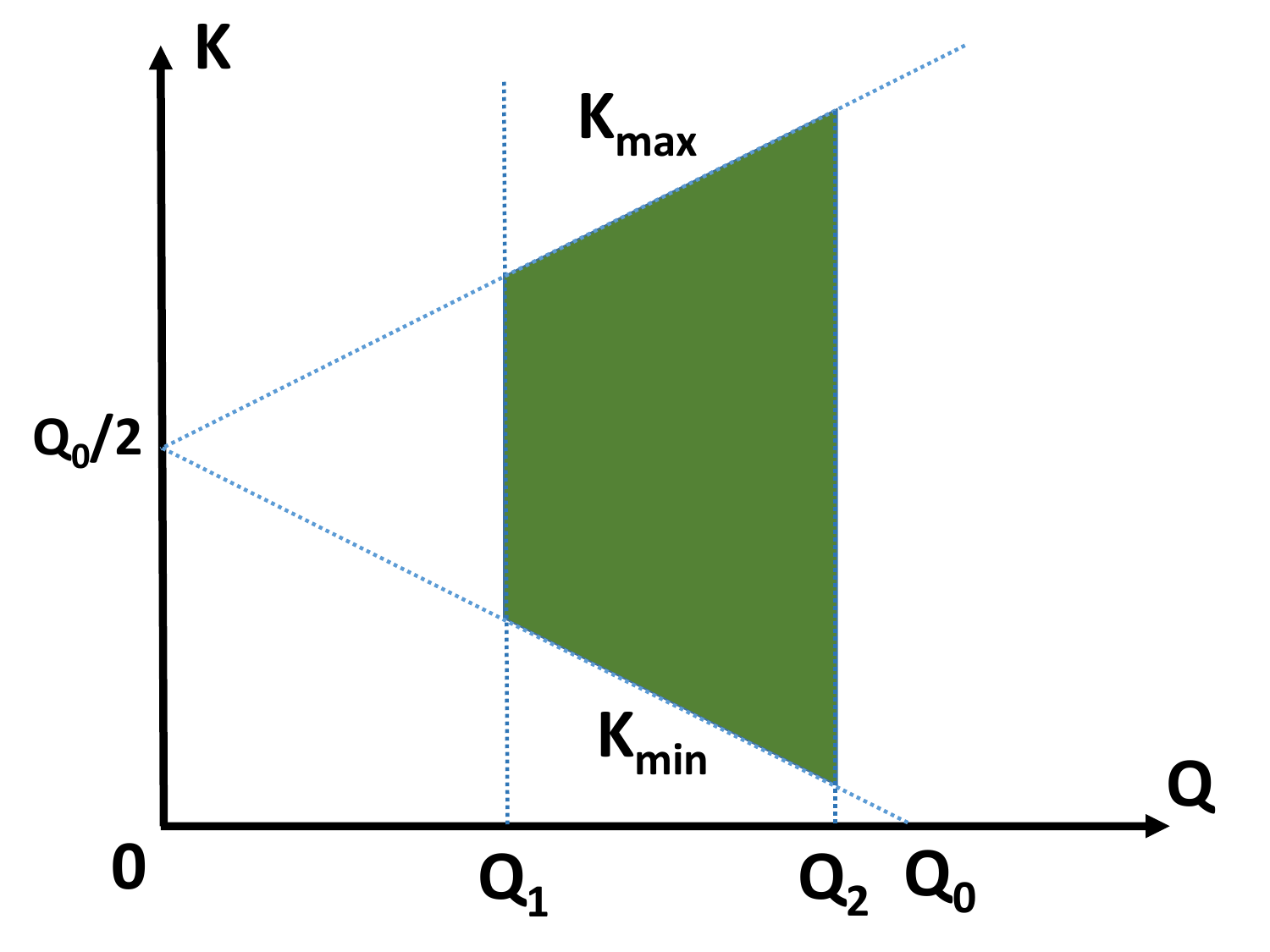}    \label{FigQKplotB}}         
    \caption{$Q-K$ integration regions (green trapezoid areas) for $(E_2-m_2,T)$ point in Dalitz region IN (left-hand side) and
                   in region OUT (right-hand side).}
    \label{FigQKplots}
\end{figure}

The kinematical limits of the bremsstrahlung photon parameters  $Q$, $K$ and $\phi_k$
can be easily obtained by Eq. \ref{ConservationEquations} and Fig. \ref{FigKinematics}
The azimuthal angle $\phi_k$ has no kinematical constraints, but the other two parameters $Q$ and $K$
of the bremsstrahlung photon are limited by the allowed green trapezoid
regions presented in Fig. \ref{FigQKplots}: the  left-hand side one (with zero minimal photon energy
$K$) is for $(E_2-m_2,T)$ points in
the Dalitz region IN, and the right-hand side one is for points in the region OUT. The $Q$-limits in
Fig. \ref{FigQKplots} are
\begin{equation} \label{Qlimits}
Q_0=\Delta-E_2-T, \quad Q_1=|P_2-P_f|, \quad Q_2=P_2+P_f 
\end{equation}
(with electron momentum $P_2=|{\bf p_2}|$ and recoil particle momentum $P_f=|{\bf p_f}|$),
and the $K$-limits are
\begin{equation} \label{Klimits}
K_{min}=(Q_0-Q)/2, \quad K_{max}=(Q_0+Q)/2
\end{equation}
(see Refs. \cite{Ginsberg1966,Ginsberg1967,Gluck1986-89,TothSzegoMargaritis1986,GluckToth1990}).
$Q_1$ and $Q_2$ are connected with the geometrical decay triangle limits for (anti)parallel ${\bf p_2}$ and ${\bf p_f}$
momentum vectors, while the $K_{min}$ and $K_{max}$ limits originate from the geometrical properties of the
rotational ellipsoid in Fig. \ref{FigKinematics}.
On Fig. \ref{FigQKplotA}, the point $(Q=Q_0,\, K=0)$ represents the 3-body decay kinematics for any point in
the region IN, while the green trapezoid represents the 4-body kinematics.

In the region IN $Q_2>Q_0$, in region OUT $Q_2<Q_0$, and on the IN-OUT boundary 
R-E $Q_2=Q_0$.
On the Dalitz plot upper boundary curve R-M
and on the right-hand side lower boundary curve E-M: $Q_1=Q_0$, i.e.
the green trapezoid region here shrinks to the vertical line $Q=Q_0$, $0<K<Q_0$.

We mention that in the case of the neutrino-type radiative correction calculations 
 \cite{YokooMorita1976,FujikawaIgarashi1976,GarciaMaya1978,Garcia1978,Garcia1981,Garcia1982},
\cite{Ando2004,Gudkov2005,Gudkov2006,IvanovPitschmann2013,
IvanovPitschmannarXiv2018,IvanovHollwieser2013,IvanovHollwieser2019,Ivanov2021}
one uses 3-body kinematics to get the connection between the neutrino and recoil particle parameters, and therefore
one cannot find there any information about the region OUT of the Dalitz plot, which is present only with 
4-body bremsstrahlung photon kinematics. 
Of course, the largest change of the 4-body kinematics relative to the
3-body kinematics on the region IN is the presence of the green trapezoid region of Fig. \ref{FigQKplotA}.
Finally, we mention also the following
special subtlety: if in the neutrino-type radiative correction calculation one integrates with respect
to the neutrino direction, then an implicit integration is performed also with respect to the recoil particle energy over the whole
IN+OUT region; i.e. in that case also the OUT region is correctly included in the calculation, and the neutrino-type
radiative correction calculation results agree with the recoil-type results (apart from very small recoil-order terms).

\section{The bremsstrahlung photon integral}
 \label{SectionBRIntegral}

Let us now scrutinize how do the above described kinematics issues influence the  radiative correction
calculations of beta and semileptonic decays. The order-$\alpha$ radiative correction to any observable
quantity is the sum of virtual and bremsstrahlung corrections. Both of them have infrared (IR) divergence, therefore
only their sum, where the IR divergent terms cancel, is an experimentally meaningful quantity. 
In the case of virtual correction, off-shell virtual
photons are created and absorbed by the charged participants of the decay. Off-shell means: energy
 and momentum of the photons are independent from each other, and virtual means: the photon 
lives only for an extremely short time during the decay process, it is constrained to the very small space-time
vicinity of the beta decay process, and is not observable from outside \cite{Yennie1961,Maximon1969,Gluck2003}. 
Therefore, the virtual part of the radiative correction has the same 3-body kinematics as the 
zeroth-order decay without radiative correction (the left-hand side part of Fig. \ref{FigKinematics}
and the regions IN in Fig. \ref{FigDalitzPlots}).
Both the zeroth-order total decay rate $\rho_0$ 
and its order-$\alpha$ model-independent (MI) or outer virtual radiative correction
$\rho_V$ can be generally calculated by 2-dimensional integrations over the Dalitz region IN
of Fig. \ref{FigDalitzPlots}: 
\begin{equation} \label{DalitzIntegralsIN}
\rho_0=\int\limits_{m_2}^{E_{2m}} dE_2 \int\limits_{T_{min}(E_2)}^{T_{max}(E_2)} dT\cdot W_0(E_2,T), \quad\quad
\rho_V=\int\limits_{m_2}^{E_{2m}} dE_2 \int\limits_{T_{min}(E_2)}^{T_{max}(E_2)} dT\cdot W_V(E_2,T). 
\end{equation}
Here the zeroth-order Dalitz distribution $W_0(E_2,T)$ is proportional to the spin averaged-summed
zeroth-order beta decay amplitude squared
$0.5 \sum_{i,2,f} |{\cal M}_0|^2$ (see the approximate formula  2.11-2.12 in Ref. \cite{Gluck1997} for unpolarized case,
and App. A in Ref. \cite{GluckToth1992} for the case with 
recoil-order corrections). 
The order-$\alpha$ virtual correction $W_V(E_2,T)$ is calculated by the interference term 
$\sum |{\cal M}_0 {\cal M}_V|$
\cite{Gluck2007,Gluck2014} (see Eqs. 3.1-3.4 in Ref. \cite{Gluck1997}).

 In the case of the bremsstrahlung part, the photon is 
on-shell, i.e. its energy is equal to its momentum (with $c_{\rm light}=1$ convention), and it is, in principle,
observable; although, the smaller the photon energy, the more difficult  its detection 
\cite{Yennie1961,Maximon1969,Gluck2003}.
For the bremsstrahlung part of the radiative correction calculations, it is important to consider the
4-body decay kinematics: the larger the photon energy, the larger the deviation of the 4-body
kinematics from the 3-body one.
The bremsstrahlung part of the total decay rate $\rho_{BR}$
can be calculated by a 2-dimensional integral over the extended Dalitz region IN+OUT of Fig. \ref{FigDalitzPlots}: 
\begin{equation} \label{DalitzIntegralINOUT}
\rho_{BR}=\int\limits_{m_2}^{E_{2m}} dE_2 \int\limits_{T_{min}'(E_2)}^{T_{max}(E_2)} dT\; W_{BR}(E_2,T), 
\end{equation}
where $T_{min}'(E_2)$ is defined by $T_{min}'(E_2)=T_{min}(E_2)$ for $E_2>E_{2h}$ (to the right from point E)
and by $T_{min}'(E_2)=0$ for $E_2<E_{2h}$ (to the left from point E; see Eqs. \ref{Tmaxmin}, \ref{ERpoints}). 
The order-$\alpha$ MI (outer)  radiative correction Dalitz distribution and total decay rate (free from IR divergence)
are obtained by adding the 
virtual and bremsstrahlung corrections:
\begin{equation} \label{Wgamma}
W_\gamma(E_2,T)=W_V(E_2,T)+W_{BR}(E_2,T), \quad  \rho_\gamma=\rho_V+\rho_{BR}.
\end{equation}

In order to calculate correctly the bremsstrahlung part of the radiative correction, one has to ask: what 
are the observables for which we need the corrections, and what are the experimental constraints 
for the observables? For example, let us assume that we want to measure the $(E_2,T)$ electron and recoil energy Dalitz 
distribution (like it is planned by the Nab experiment \cite{Nab2017}).
 Then the main questions are: a, Are there any experimental constraints about the angle 
between the electron and recoil particle? b, Are the bremsstrahlung photons (and the neutrinos) 
 detected? If the answer is twice no, then the main task of the order-$\alpha$ bremsstrahlung correction 
calculation (without any polarization information) is to perform the following 3 dimensional integral:
\begin{equation} \label{WBR}
W_{BR}(E_2,T)=\frac{1}{2^{10} \pi^6 m_i}
\int\limits_{Q_1}^{Q_{\rm max}} dQ
\int\limits_{K_{\rm min}}^{K_{\rm max}} dK  \frac{K}{\sqrt{K^2+m_\gamma^2}}
\int\limits_{0}^{2\pi} d\phi_k \, {\bar M}_{BR},
\end{equation}
where $Q_{\rm max}=Q_0$ in region IN and $Q_{\rm max}=Q_2$ in region OUT (see Figs. \ref{FigDalitzPlots}
and \ref{FigQKplots}),
$m_\gamma$ denotes the photon mass (which is used for the infrared regularization), and
${\bar M}_{BR}= (1/2)\sum_{i,2,f,k} |{\cal M}_{\rm BR}|^2$ 
(depending on $E_2$, $T$, $Q$, $K$ and $\phi_k$)
denotes the
bremsstrahlung transition amplitude squared averaged over the decaying particle $i$ polarization and
summed over the polarization states of the electron, recoil
particle and photon; see App. \ref{SectionDerivationRecoilType} for the derivation of Eq. \ref{WBR}.
The computation of the  bremsstrahlung transition amplitude ${\cal M}_{BR}$, containing
Dirac matrix traces and Lorentz index contractions,
can be evaluated either by hand on paper (in the simplest cases), or
by computer symbolic algebra (like REDUCE or Mathematica);
see Refs. \cite{TothSzegoMargaritis1986,Gluck1986-89,GluckToth1990,GluckToth1992}
about details of these calculations.
In the case of polarized neutron decay (without summing over the decaying neutron polarization),
our $\sum_{2,f,k} |{\cal M}_{\rm BR}|^2$
formula (\cite{GluckGENDER}, without recoil-order terms)
agrees with Eq. B10 in Ref. \cite{IvanovPitschmannarXiv2018}
(in Eq. 2.36 of Ref. \cite{CooperPhD} the coefficients of $A$ have wrong sign).
The unpolarized formula $\sum_{i,2,f,k} |{\cal M}_{\rm BR}|^2$
can be found in the appendix of Ref. \cite{Gluck1993}, in Eqs. 4.4-4.7 of Ref. \cite{Gluck1997},
or in Eq. 2.29 in Ref. \cite{CooperPhD}.

All  3 integrals in Eq. \ref{WBR} can be expressed analytically, although one obtains many 
and rather complicated formulas (but very fast computer codes for the evaluation of these formulas); see Refs. 
\cite{tHooftVeltman,Nagy,Gluck1986-89,GluckToth1990}
about the 3-dimensional analytical integration method used in our publications.
The 3-dimensional integral can also be performed completely numerically 
\cite{TothMargaritisSzego1984,TothSzegoMargaritis1986,ChristianKuhnelt1978}.
Semianalytical  integration methods for the non-infrared divergent
 integrals, with analytical integrals over the parameters $\phi_k$ and $K$ and 
numerical integrations over $Q$, are described in Refs.
\cite{Gluck1986-89,GluckToth1992,Gluck1993}.  
Recently, we developped the C++ class SANDI 
(Semi-Analytical Neutron Decay Integrator) \cite{GluckSANDI} which is based on this
semianalytical  integration method; all  recoil-type radiative correction results 
presented in the following Sec.
\ref{SectionComparison} were computed by using this code,
while all our previously published radiative correction results were computed by 
using the FORTRAN programming language
(our  new radiative correction results with the SANDI code agree with the old results
published in Refs. \cite{Gluck1993,Gluck1998a}).
The above described analytical and semianalytical methods are technically quite different, but all our
bremsstrahlung integral results of type \ref{WBR} computed by these two methods agree with 13 digits;
this is an important testing possibility in order to exclude various kinds of computation errors.

We performed several comparisons of our bremsstrahlung  integrals with published results: a, with Eqs. 27 and
A1-A4 in Ref. \cite{Ginsberg1967}; with App. C of Ref. \cite{Cirigliano2004}; and with Eqs. D9-D12 of Ref. \cite{Seng2021b}
(the latter formula  is especially useful, because it is rather simple and applicable for both charged and neutral hadron decays).
In addition, all our bremsstrahlung analytical integral results (even the infrared divergent ones) were compared 
also by numerical integrations.
In all these comparisons, complete (10 or more digits) agreement was found.
We mentioned above that for the regularization of the infrared divergent integrals (with the order $1/K^2$ terms
in ${\bar M}_{BR}$) we use the photon mass method (one gets the same results with dimensional regularization
\cite{MarcianoSirlin1975,Seng2021b}). The infrared divergent integrals are independent of whether one uses
 the $Q$ ad $K$ integration limits with zero photon mass (Eqs. \ref{Qlimits} and \ref{Klimits}) or with the same finite photon mass
as in ${\bar M}_{BR}$ (Eq. \ref{KminKmax}).

We mentioned in Sec. \ref{SectionIntroduction} that there are many publications for radiative correction calculations
of hyperon and meson semileptonic decays (Refs.
 \cite{Ginsberg1966,Ginsberg1967,Ginsberg1968,Ginsberg1970,Becherrawy1970,KamalWong1971,Tun1989,Tun1991,Martinez1993,Juarez1993,Martinez2000,FloresMendieta2006, 
Cirigliano2002,Bytev2003,Cirigliano2004,Andre2005,Andre2007,JuarezLeon2011,Neri2015,Seng2021a,Seng2021b,Seng2022}).
Most of these publications use also analytical or semianalytical integration methods for the bremsstrahlung correction
calculations. The integration variables used by these papers are different from our variables $Q$, $K$ and $\phi_k$;
the analytical results of these publications, similarly to our results, are rather complicated.
All the analytical and semianalytical bremsstrahlung integrations available in the literature are
rather difficult, and different observable quantities require the calculations of different
integrals. On the other hand,
the Monte Carlo integration method (see Refs. \cite{GluckJoo1996,Gluck1997,GluckJoo1997}
and references therein) provides a much simpler possibility to compute Eq. \ref{WBR} and similar multidimensional 
bremsstrahlung integrals.
Using  the Monte Carlo method, the computer takes over a large part of the integration difficulties, resulting in a significant reduction of
the human efforts (there is no need to go through many complicated analytical integrals). 
The accuracy is limited by statistics, but
there is no problem to generate a few hundred million events within a reasonable computation time (few minutes),
and this provides already meaningful radiative correction results. The MC method is very flexible, since
the same computer code can be used in order to calculate
radiative corrections to any kind of measurable quantity; and it is
especially suitable for experimental off-line data analyses, where the
various kinematic cuts, detection efficiencies etc. require complicated
modifications of the theoretical distributions (a good example is the radiative correction calculation for
the aCORN experiment \cite{aCORN2021}).
Using the Monte Carlo method
one can perform both weighted or unweighted event generations.
Our Monte Carlo codes of  \cite{GluckJoo1996,Gluck1997,GluckJoo1997}
were written in the nineties with FORTRAN. Recently, we have written a new
C++ Monte Carlo class called GENDER: GEneration of polarized Neutron (and nuclear beta) Decay Events with
Radiative and recoil corrections  \cite{GluckGENDER}. This code uses an improved version of the
Monte Carlo method described in Ref. \cite{Gluck1997}. Many of our test computations show very good agreement
between the semianalytical SANDI and the Monte Carlo GENDER radiative correction results
(e.g. with difference of 0.00001\% for the relative correction of the proton spectrum that is integrated above 400 eV, by using 10 billion MC events),
and we made also
many test comparisons of the SANDI and GENDER calculations with published results \cite{GluckSANDI,GluckGENDER}.
We mention that the Monte Carlo method has been used in the past few decades for many theoretical computations
in high energy physics; unfortunately, this powerful method seems to be not so 
popular among the physicists involved in radiative correction calculations of beta decays.  
An exception is by Refs. \cite{Andre2005,Andre2007} where the author uses  Monte Carlo integration methods.

We would like to understand now:
why is the analytical integration of Eq. \ref{WBR} so difficult? The main reason is the fixed recoil particle kinetic energy $T$:
in this case the recoil particle (proton) momentum magnitude 
$P_f=|{\bf p_f}|=\sqrt{2m_fT+T^2}$ is also fixed, and this causes
strong constraints for the neutrino and bremsstrahlung photon momenta. Namely,
due to momentum conservation in the decaying particle CMS:
$P_f=|{\bf p_2}+{\bf p_1}+{\bf k}|$.
Therefore, the ${\bf p_2}$, ${\bf p_1}$ and ${\bf k}$ momentum vectors are strongly constrained by this equation,
and the resulting integration
limit correlations make the analytical integrations cumbersome;
Eqs. \ref{integralnext} and \ref{dcosthetak} in App. \ref{SectionDerivationRecoilType}
illustrate the  correlations among the various integration variables.
This complication of the kinematical limits is present even in the case of the proton (recoil particle) energy
spectrum calculation, where one integrates over the electron energy $E_2$.
Namely, $E_2$ is connected to the neutrino energy $E_1$ and photon energy $K$ by energy
conservation, and therefore the electron momentum ${\bf p_2}$ is not able to fulfill the momentum conservation
equation ${\bf p_f}+{\bf p_2}+{\bf p_1}+{\bf k}=0$ for arbitrary neutrino and photon momenta.

The situation is different in the case of the electron energy spectrum calculation, where the recoil particle kinetic energy
is not fixed but is integrated over all possible values; especially for small $\Delta$ values (like it is the case
for neutron and nuclear beta decays), 
when the 
recoil particle kinetic energy is small. In this case, the recoil particle kinetic energy $T$ can be neglected
in the energy conservation equation: $\Delta\approx E_1+E_2+K$, and therefore the
momentum ${\bf p_f}$ can freely fulfill the momentum 
conservation constraint ${\bf p_f}+{\bf p_2}+{\bf p_1}+{\bf k}=0$ for arbitrary electron, 
neutrino and photon momenta (see the bremsstrahlung integral derivation in App. \ref{SectionDerivationNeutrinoType}).
One can see in the bremsstrahlung integral of Eq. \ref{integralneutrinotype} in App. \ref{SectionDerivationNeutrinoType}
and of Eq. 4.10 in Ref. \cite{Gluck1997} that, after the elimination of the recoil particle momentum
(and neglecting the recoil-order terms),
 the electron, neutrino and photon directions 
have no kinematical constraints.
Due to the simple integration limits, one obtains uncomplicated analytical formulas for the order-$\alpha$
model independent radiative corrections to the electron energy spectrum \cite{Sirlin1967}
and to the electron asymmetry 
(angular correlation of the electron momentum to the decaying particle spin) \cite{Shann1971}. 
We mention that the large proton/electron mass ratio causes a large asymmetry 
in the electron and proton spectrum radiative calculations
not only at kinematics, but also at dynamics: one can see in Eqs. 4.4-4.8 of Ref. \cite{Gluck1997}
and Eqs. A.7-A.13 of Ref. \cite{Gluck1993} that the bremsstrahlung amplitude squared has strong dependence on the
electron and neutrino momenta and energies, but almost no dependence on the proton (recoil particle) momentum
 and energy (the light electron can much easier radiate a bremsstrahlung photon than the heavy proton).

Several authors observed in the seventies 
\cite{YokooMorita1976,FujikawaIgarashi1976,GarciaMaya1978}
that the  bremsstrahlung photon integrations remain simple even if one fixes, in addition to the electron energy
and direction, also the neutrino direction vector (but not the neutrino energy): similarly to the case
 explained in the previous paragraph,
the recoil particle momentum can easily fulfill the momentum conservation equation, and so
the bremsstrahlung integration limits are rather simple (see App. \ref{SectionDerivationNeutrinoType}).
After including the virtual corrections,
the order-$\alpha$ outer (model-independent) radiatively corrected energy and angular distribution
of polarized neutron decay (without recoil-order corrections) can be then written with the following simple formula
\cite{GarciaMaya1978,Garcia1981,Ando2004,Gudkov2006,IvanovPitschmann2013}, as a radiative correction
generalization of the well-known zeroth-order formula of Jackson, Treiman and Wyld
 \cite{Jackson1957}:
\begin{equation} \label{garciamaya}
w_{0\gamma}(E_2,c_1,c_2, c_{12})=\frac{w_{e0}(E_2)}{16\pi^2} (G(E_2)+H(E_2) a\beta c_{12}+
H(E_2) P A\beta c_2 +G(E_2) P B c_1),
\end{equation}
\begin{equation} \label{GH}
G=1+\frac{\alpha}{2\pi} g(E_2), \quad H=1+\frac{\alpha}{2\pi} h(E_2),
\end{equation}
\begin{equation} \label{wint}
\rho_{0\gamma}=\rho_0+\rho_V+\rho_{BR}=\int dE_2 \int d\Omega_2  \int d\Omega_1\cdot w_{0\gamma}(E_2,c_1,c_2, c_{12}),
\end{equation}
where $w_{e0}(E_2)$ is the zeroth-order electron energy spectrum,
$c_2$, $c_1$ and $c_{12}$ denote the cosines of the electron and neutrino polar angles relative
to the neutron spin and the cosine of the electron-neutrino angle, respectively; $P$ is the neutron polarization, 
$\beta=P_2/E_2$ the electron velocity, and
$a$, $A$ and $B$ are the well-known
electron-neutrino correlation parameter, and the electron and neutrino asymmetry parameters, respectively.
 The analytical radiative correction
functions $g(E_2)$ and $h(E_2)$ (see App. \ref{SectionRadCorrFormulas})    are defined originally by Sirlin in Eq. 20b of Ref. \cite{Sirlin1967} and 
by Shann in Eq. 10 of Ref. 
\cite{Shann1971}; nevertheless, they appear (with other notations) in many later 
publications.\footnote{$g=g(E_2)=2(\phi_1+\theta_1)$ and $h=h(E_2)=2(\phi_2+\theta_2)$ of Eqs. 
15 and 16 in Ref. \cite{GarciaMaya1978}; $g=2g_n$ and $h=2(g_n+f_n)$ of Eq. A7 in \cite{IvanovPitschmann2013}
(without the model-dependent correction term $C_{WZ}$);
$h-g=\delta_\alpha^{(2)}$ of Eq. 12 in \cite{Ando2004}, etc.} 
We mention that we were able to test  our Monte Carlo code GENDER
\cite{GluckGENDER,Gluck1997}
by comparison with Eqs. \ref{garciamaya}, \ref{GH} for polarized neutron decay, 
and we found very good agreement between these two 
completely different (analytical and Monte Carlo integration) computation methods. 
This fact shows also that Eqs. \ref{garciamaya}, \ref{GH} are correct and, \textbf{in principle}, could
be applied as radiative correction results for (hypothetical) polarized neutron decay experiments
where \textbf{the neutrino direction is observed by explicit neutrino detection}.
The authors of Refs. \cite{YokooMorita1976,FujikawaIgarashi1976,GarciaMaya1978,Garcia1981,Garcia1982}
know of course that this is experimentally unrealistic. Instead, they
claim that, using the 3-body decay kinematics relations among the electron, neutrino and recoil
particle, one can use Eqs. \ref{garciamaya} and \ref{GH} as  radiative correction calculations for the
analyses of experiments which
detect both the electrons and the recoil particle (e.g. proton in neutron decay), with the purpose
of electron-neutrino correlation or neutrino asymmetry parameter determinations.
Many other publications in the past 2 decades
\cite{Ando2004,Gudkov2005,Gudkov2006,IvanovPitschmann2013,
IvanovPitschmannarXiv2018,IvanovHollwieser2013,IvanovHollwieser2019,Ivanov2021}
use Eqs.  \ref{garciamaya} and \ref{GH} as simple and elegant radiative correction results for
neutron decays.

Nevertheless, just the abovementioned energy and momentum conservation properties, 
which are beneficial for the analytical integrations, make these neutrino-type radiative correction results
inappropriate for the precision analyses of beta decay experiments with recoil particle detection. 
Namely, due to the momentum conservation in the decaying particle CMS, 
with fixed electron momentum ${\bf p_2}$ and neutrino direction unit vector $\hat{\bf p}_1$,
and with neutrino energy $E_1\approx \Delta-E_2-K$,
the recoil particle momentum in the decaying particle CMS 
${\bf p_f}=-{\bf p_2}-E_1 \hat{\bf p}_1-{\bf k}$  (and thus also the recoil particle kinetic energy $T$)
 is  integrated during the bremsstrahlung phase space integration,
together (in strong correlation) with the photon momentum
${\bf k}$. Therefore  all useful information about the 
energy or direction of the recoil particle, which would be important for the radiative correction
calculation results,
 is lost after the bremsstrahlung photon integration; in fact, also
the neutrino energy $E_1$ information is lost (this information loss is illustrated by an example in the next paragraph).
The information loss would be not present if the bremsstrahlung photon energy $K$ were
limited to small ($K<<\Delta-E_2$) values; but this is usually not the case, since
$K_{\rm max}\approx \Delta-E_2$ (see Eq. \ref{integralneutrinotype}).
Due to to this information loss, it is not possible to determine the neutrino direction
by using the measured electron and recoil particle properties; at least not at the 
radiative correction calculation precision level. 
Therefore the neutrino-type radiative correction results of Eqs. \ref{garciamaya} and \ref{GH}
have only academic interest, and they are not suitable for neutron or nuclear
beta decay experimental analyses where the proton (recoil particle) is detected,
with the purpose to determine the electron-neutrino correlation parameter $a$
\cite{aCORN2021,aSPECT2020,Nab2017}
or the neutrino asymmetry parameter $B$ \cite{Schumann2007}.

\begin{figure}[htbp]
    \centering
    \subfigure[proton energy distribution]{\includegraphics[width=0.47\textwidth]
                  {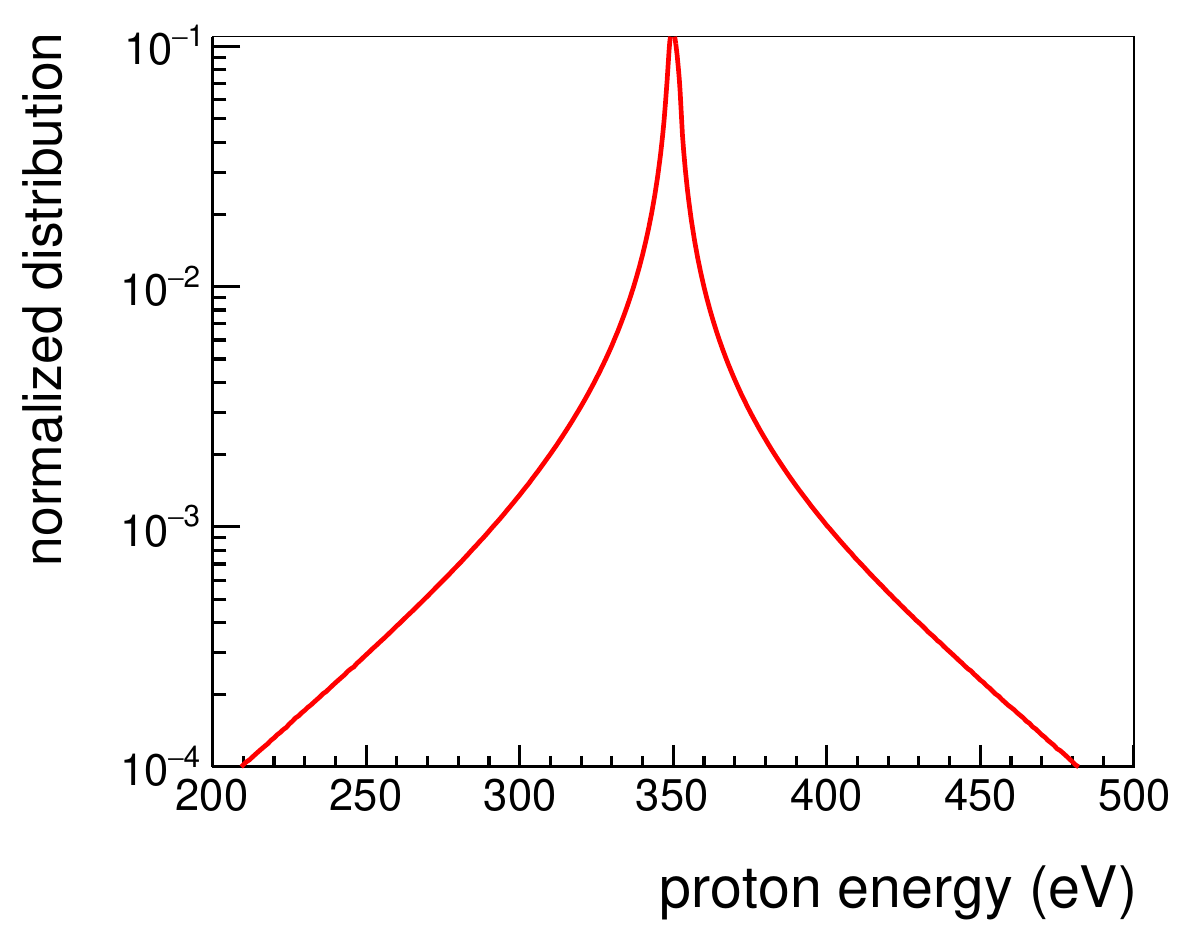}  \label{FigT1}}\quad
    \subfigure[neutrino energy distribution]{\includegraphics[width=0.47\textwidth]
                  {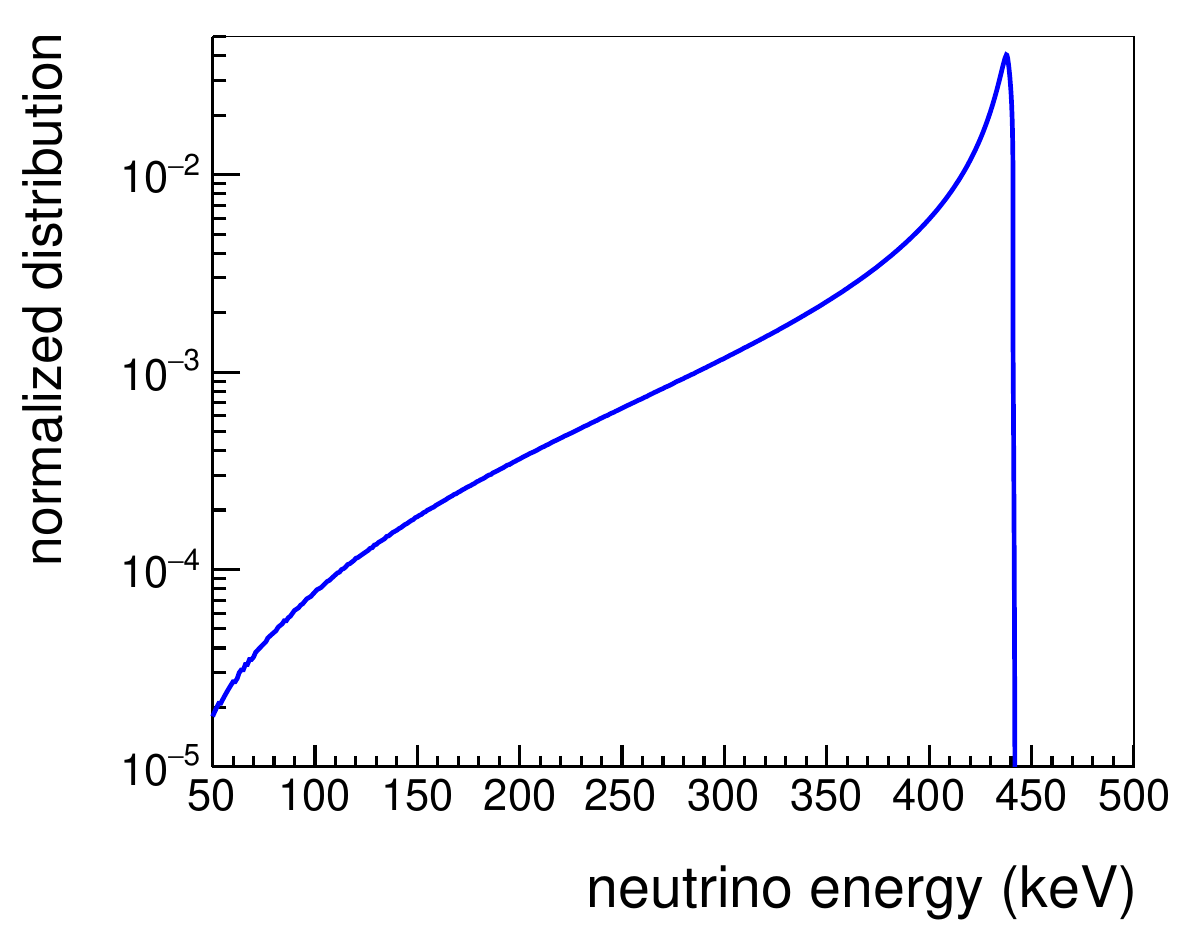}    \label{FigE1}}         
    \caption{Proton and neutrino  energy ($T$ and $E_1$) distributions in neutron decay with fixed $E_2-m_2=340$ keV electron kinetic
          energy and $90^\circ$ electron-neutrino angle. With $K=0$ bremsstrahlung photon energy: $T=T_0=350.87$ eV, 
           $E_1=E_{10}=441.98$ keV. Only the $|T-T_0|>1$ eV events (with 0.57 \% probability) are plotted.                }
    \label{FigTE1}
\end{figure}

Fig. \ref{FigTE1} illustrates the above explained information loss with the following unpolarized neutron decay example, with fixed
electron kinetic energy $E_2-m_2=340$ keV and an angle $90^\circ$ between
electron and neutrino momentum directions. 
In zeroth-order
(without radiative corrections), the proton and neutrino have in this case fixed kinetic energies $T=T_0=350.87$ eV
and $E_1=E_{10}=441.98$ keV, respectively. In the presence of bremsstrahlung photons the proton and neutrino 
are correlated with the photon, and therefore their
energies have wide
distributions, which are peaked around the zeroth-order values $T_0$ and $E_{10}$, as one can see in Fig. \ref{FigTE1} 
(the distributions were computed by using the unweighted event generator member function of
the C++ code GENDER \cite{GluckGENDER}; 
the normalization is defined by setting the sum of the 1 eV 
binning function values equal to 1). The proton and the neutrino energy are connected to the bremsstrahlung photon 
mainly by momentum and energy conservation, respectively; this is why the two distributions have different behaviour
(the proton energy can be either smaller or larger than the zeroth-order value $T_0$, 
while the neutrino energy is always smaller than $E_{10}$).
The large $|T-T_0|$ events have rather small probabilities (e.g. 0.21 \% and 0.09 \% probability 
for $|T-T_0|>10$ eV and $|T-T_0|>30$ eV, respectively). Nevertheless, even these small probability events can invalidate
the precise kinematic connection between the recoil particle parameters and the neutrino direction, which is used
by the neutrino-type radiative correction calculations. 

We show in the following section that the neutrino-type radiative correction results are usually 
(for observables with recoil particle detection) completely different
from the  recoil-type results; these differences are obviously caused by the incorrect kinematic connection
of the neutrino-type calculations.
We mention that the recoil-type and neutrino-type
radiative correction calculations agree 
(neglecting very small recoil-order terms) only for
those observables where the recoil particle is not detected, and thus not used in the experimental analysis 
(like the electron energy spectrum or the electron asymmetry).

We close this section by stating that, in  addition to  Eq. \ref{garciamaya},
there are other radiative correction observables where
one can present the results by simple analytical formulas: a, the classical KUB formula with the
bremsstrahlung photon energy spectrum,
in coincidence with the electron energy (see Sec. 4 in  Ref. \cite{Gluck1997}, with references therein);
b, radiative correction to the neutrino energy spectrum \cite{BatkinSundaresan1995,Sirlin2011,SirlinFerroglia2013}
\footnote{Replace in Eq. 12 of Ref. \cite{BatkinSundaresan1995} $-\beta(0)$  by -1, and the
last row of Eq. 14 by $-3\alpha/(8\pi)$; then g of Eq. 14 in \cite{BatkinSundaresan1995} 
is identical with $\alpha/(2\pi) h$ of Eq. 11 in Ref. \cite{Sirlin2011}
(the latter agrees also with our own semianalytical calculation).}
(Eq. 4.12 in Ref. \cite{Gluck1997}, with 4.9, can also be used as neutrino and photon double energy distribution).
In all these observables (like in Eq. \ref{garciamaya}), the recoil particle energy (momentum) is absent,
i.e. it is completely integrated over the bremsstrahlung phase space; this enables
the relatively simple analytical bremsstrahlung integrations.
Our Monte Carlo computations \cite{Gluck1997,GluckGENDER} have very good agreement also with these analytical results.
We mention that, similarly to Eq. \ref{garciamaya}, the analytical radiative correction formulas 
to the neutrino energy spectrum  of Refs. \cite{BatkinSundaresan1995,Sirlin2011,SirlinFerroglia2013}
are applicable only to beta decay experiments with explicit neutrino detection, and not for experimental analyses where
the neutrino energy is determined from electron and recoil particle energy by zeroth-order 3-body kinematics.

\section{Recoil-type and neutrino-type outer radiative corrections}
 \label{SectionComparison}

In this section, after a presentation of the electron spectrum radiative corrections for two decays,
we compare the recoil-type and neutrino-type outer radiative corrections for 3 distributions
of unpolarized neutron and nuclear beta decay: the 2-dimensional $(E_2,T)$ and $(E_2,\cos \theta_{e\nu})$ Dalitz
distributions, and the 1-dimensional proton energy spectrum. The recoil-type calculation results presented here were computed
with our new C++ code SANDI \cite{GluckSANDI}.

\begin{figure}[htbp]
    \centering
    \subfigure[neutron decay]{\includegraphics[width=0.47\textwidth]
                  {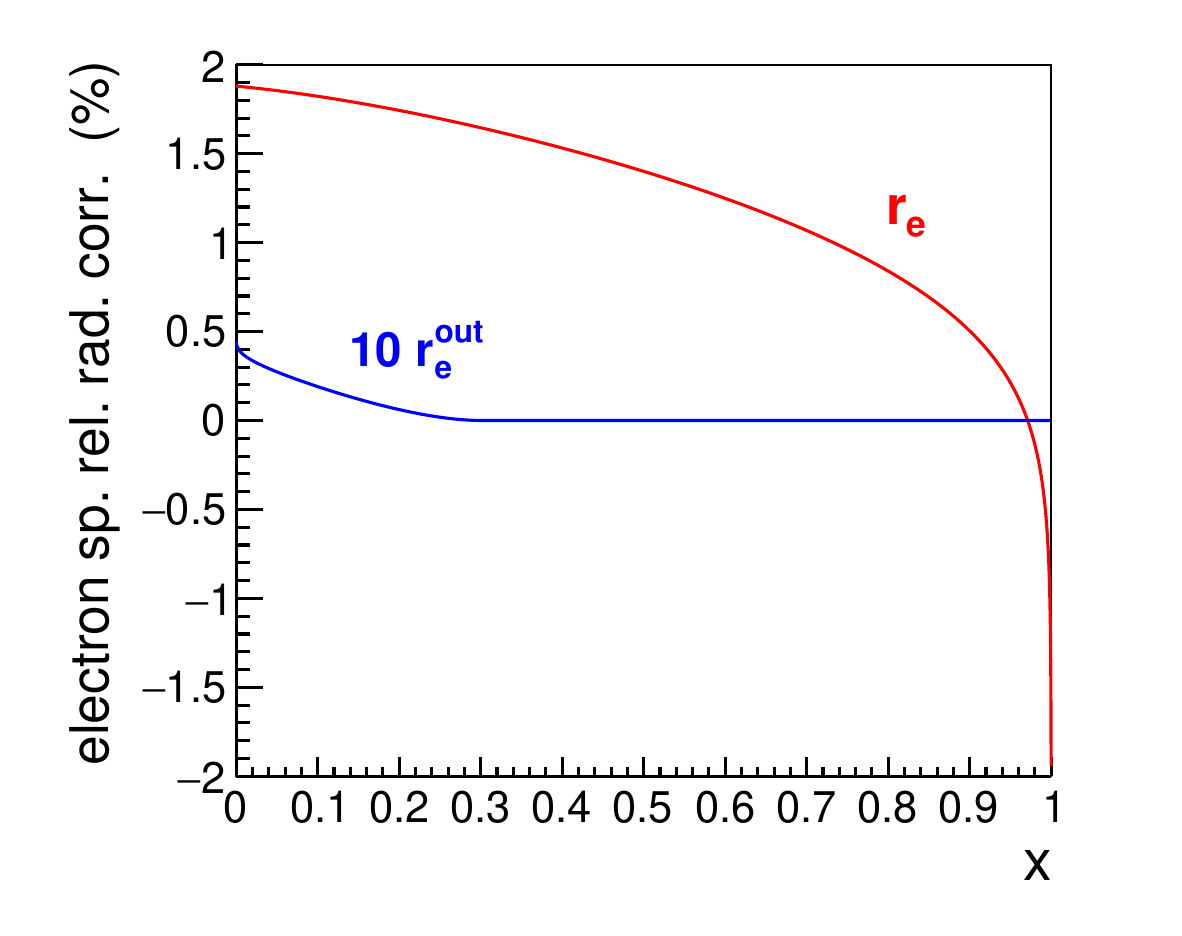}  \label{FigElectron1}}\quad
    \subfigure[10 MeV nuclear decay]{\includegraphics[width=0.47\textwidth]
                  {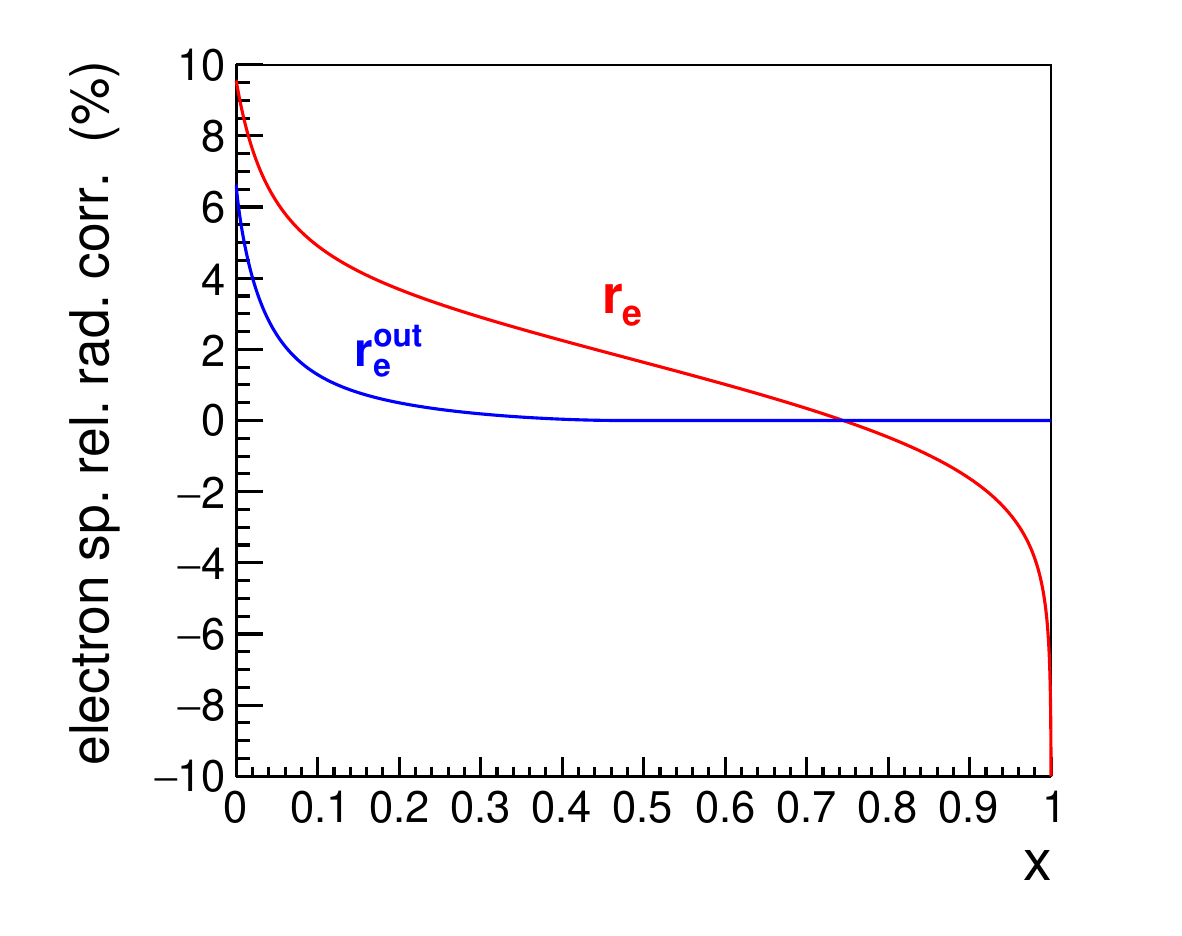}    \label{FigElectronNB}}         
    \caption{Relative radiative corrections $r_e$ (red curves) and $10\, r_e^{out}$ or $r_e^{out}$ (blue curves)   
                 to the electron energy spectrum for neutron decay (left), 
                and nuclear $\beta^-$ decay with $\Delta=10$ MeV and $m_f=20\; m_{\rm proton}$ (right).
                In this case there is practically no difference (less than 0.001 \%, due to
              small differences in recoil-order terms) between the recoil-type and neutrino-type
                calculations (the recoil-type calculation was used for this plot).                }
    \label{FigElectron}
\end{figure}

In Fig. \ref{FigElectron} we present the order-$\alpha$ MI (outer) radiative correction $r_e$ (red curves) 
and  $r_e^{out}$ (blue curves)  to the electron energy spectrum in neutron decay
and in a nuclear $\beta{}^-$-decay with $\Delta=10$ MeV and $m_f=20\; m_{\rm proton}$. 
We determine in this section the electron energy $E_2$ (and later below the recoil particle energy $T$)
with the dimensionless parameters $x$, $y$ and $y_p$ ($0<x<1$, $0<y<1$, $0<y_p<1$) as
\begin{equation} \label{xy}
E_2=m_2+(E_{2m}-m_2)x, \quad T=T_{min}(E_2)+(T_{max}(E_2)-T_{min}(E_2))y=y_p T_m,
\end{equation}
where the Dalitz-plot boundary functions $T_{min}(E_2)$ and $T_{max}(E_2)$, and the
maximum recoil particle energy $T_m$,  are described in Eqs. \ref{Tmaxmin} and \ref{Mpoint},
and plotted in Fig. \ref{FigDalitzPlots}.
Our definition of the relative corrections $r_e=r_e(x)$ and $r_e^{out}=r_e^{out}(x)$ is
\begin{equation} \label{Eqre}
r_e=100\left(\int\limits_{T_{min}'(E_2)}^{T_{max}(E_2)} dT \cdot W_\gamma(E_2,T)\right)/w_{e0}(E_2),\quad
r_e^{out}=100\left(  {\int\limits_{T_{min}'(E_2)}^{T_{min}(E_2)} dT\cdot W_{BR}(E_2,T)},\right)/w_{e0}(E_2)
\end{equation}
where the generalized boundary $T_{min}'(E_2)$ is defined after Eq. \ref{DalitzIntegralINOUT}, and
the zeroth-order electron spectrum $w_{e0}(E_2)$ is the integral of $W_0(E_2,T)$
with respect to $T$. The correction $r_e^{out}$, defined by the bremsstrahlung correction integral over the region
OUT, is not presented in the neutrino-type radiative correction papers, because the simple analytical integration
of the electron spectrum radiative correction calculation
works only over the joined IN+OUT region, but not separately in region IN or OUT.
As we emphasized earlier, in the case of the neutrino-type radiative correction calculations one integrates completely over the recoil
particle phase space, therefore all information about the recoil particle (which is necessary for the distinction of these
two regions) is lost.
  In the electron spectrum case, practically (except of small recoil-order terms) there is no
difference between the recoil-type $r_e$ and neutrino-type radiative correction results, because the recoil particle
is not observed. The neutrino-type correction in this case is proportional to  the Sirlin function $g(E_2)$
(see Secs. \ref{SectionIntroduction},  \ref{SectionBRIntegral} and appendix \ref{SectionRadCorrFormulas}).
The relative difference of our $r_e$ results and of  the Sirlin function correction $r_{e,Si}=100(\alpha/(2\pi)) g(E_2)$
is about $(r_e-r_{e,Si})/r_e \sim 10^{-4}$. The small deviation between the recoil-type and neutrino-type
electron spectrum radiative correction results is due to differences in the small recoil-order terms.

It is conspicuous in Fig. \ref{FigElectron} that the correction function $r_e(x)$ has a logarithmic singularity near the electron energy
maximum $E_2=E_{2m},\; x=1$. It can be generally written in the form:
\begin{equation} \label{regen}
r_e(x)=100(f_1(x)+f_2(x)\ln (1-x)), \quad f_2(x)=\frac{\alpha}{\pi}\left[\frac{1}{\beta}\ln 
\left(\frac{1+\beta}{1-\beta}\right)-2\right],
\end{equation}
where the function $f_1(x)$ (determined by Sirlin's universal correction function $g(E_2)$ presented in Eq. \ref{gE2})
has no singularity at $x=1$; $\beta=P_2/E_2$ denotes the electron velocity.
The singularity in Eq. \ref{regen}
is due to the vanishing
bremsstrahlung photon phase space volume in the $x\to 1$ limit,
 and thus due to the appearance of the infrared divergence of the
virtual correction. This singularity can be avoided by exponentiation 
(see Refs. \cite{Maximon1969,YennieSuura1957,Yennie1961,Matsson1969,RoosSirlin1971,RepkoWu1983}):
\begin{equation} \label{exponentiation}
1+f_2(x)\ln (1-x) \Rightarrow e^{f_2(x)\ln (1-x)} =(1-x)^{f_2(x)}.
\end{equation}
Nevertheless, the above described logarithmic singularity causes
no practical problem  in the experimental analyses, therefore the exponentiation is usually not necessary
(e.g. to get  $1+f_2(x)\ln (1-x)=0.9$  in neutron decay, the parameter  $1-x$ has to be $10^{-15}$).

In addition to the logarithmic singularity near the end point, one can also notice in Fig. \ref{FigElectron}
that the radiative correction to the electron energy spectrum increases with the decay energy 
parameter $\Delta$ (the correction is much larger at the right-hand side);
see also Fig. \ref{FigDalitz} below, and compare the neutron decay \cite{Gluck1993} 
and hyperon semileptonic decay \cite{GluckToth1990}
radiative correction results.
This behaviour is connected with the KLN-theorem 
\cite{KinoshitaSirlin1959,Kinoshita1962,LeeNauenberg1964,Sirlin2011,SirlinFerroglia2013}:
due to collinear peaks of the Feynman amplitudes (when the photon momentum is nearly 
parallel to the electron momentum), the radiative corrections to the quantities with
non-integrated electron energy contain logarithmic $\ln(E_2/m_2)$-type terms, which are singular
in the $m_2\to 0$ limit. On the other hand, the radiative corrections to the observables
with integrated electron energy (like the total decay rate, the recoil particle or the neutrino energy spectrum)
do not have this strong $\Delta$-dependence and  are finite in the $m_2\to 0$ limit.

Fig. \ref{FigDalitz} shows the recoil-type and neutrino-type relative radiative corrections 
to the $(E_2,T)$ Dalitz distribution of neutron decay (left) and 
of nuclear beta${}^-$ decay with $\Delta=10$ MeV and $m_f=20\; m_p$ (right)
for
two different electron energies; in the case of the hypothetical nuclear decay,
the neutron decay coupling constants are used for the calculations.
For fixed electron energy, the proton kinetic energy $T$ is defined by the dimensionless parameter $y$ (see Eq. \ref{xy}).
The relative radiative correction $r$ of the Dalitz distribution $(E_2,T)$
shown on the vertical axes of 
Fig. \ref{FigDalitz} is defined as
\begin{equation} \label{Eqr}
r=r(x,y)=100\frac{W_\gamma(E_2,T)}{W_0(E_2,T)}-r_e(x),
\end{equation}
where $W_0(E_2,T)$ denotes the zeroth-order Dalitz distribution, $W_\gamma(E_2,T)$ is the
order-$\alpha$ model independent (outer) radiative correction part of this distribution (see Eq. \ref{Wgamma}),
and the electron spectrum correction $r_e(x)$ was defined above. The $y$-independent correction $r_e(x)$
is subtracted here in order to get smaller  values for $r$, and thus to see better the difference between
the neutrino-type and recoil-type corrections; another advantage is that due to this subtraction the correction $r$ does
not change by adding a $T$-independent term to the relative Dalitz correction  $W_\gamma(E_2,T)/W_0(E_2,T)$.
We computed the neutrino-type radiative correction by 3-body kinematics transformation from Eqs. 
\ref{garciamaya}, \ref{GH} (see Eq. A1 in Ref. \cite{Gluck1998a}, and Eqs. I-6 to I-15
 in Ref. \cite{IvanovPitschmannarXiv2018}).

\begin{figure}[htbp]
    \centering
    \subfigure[neutron decay]{\includegraphics[width=0.47\textwidth]
                  {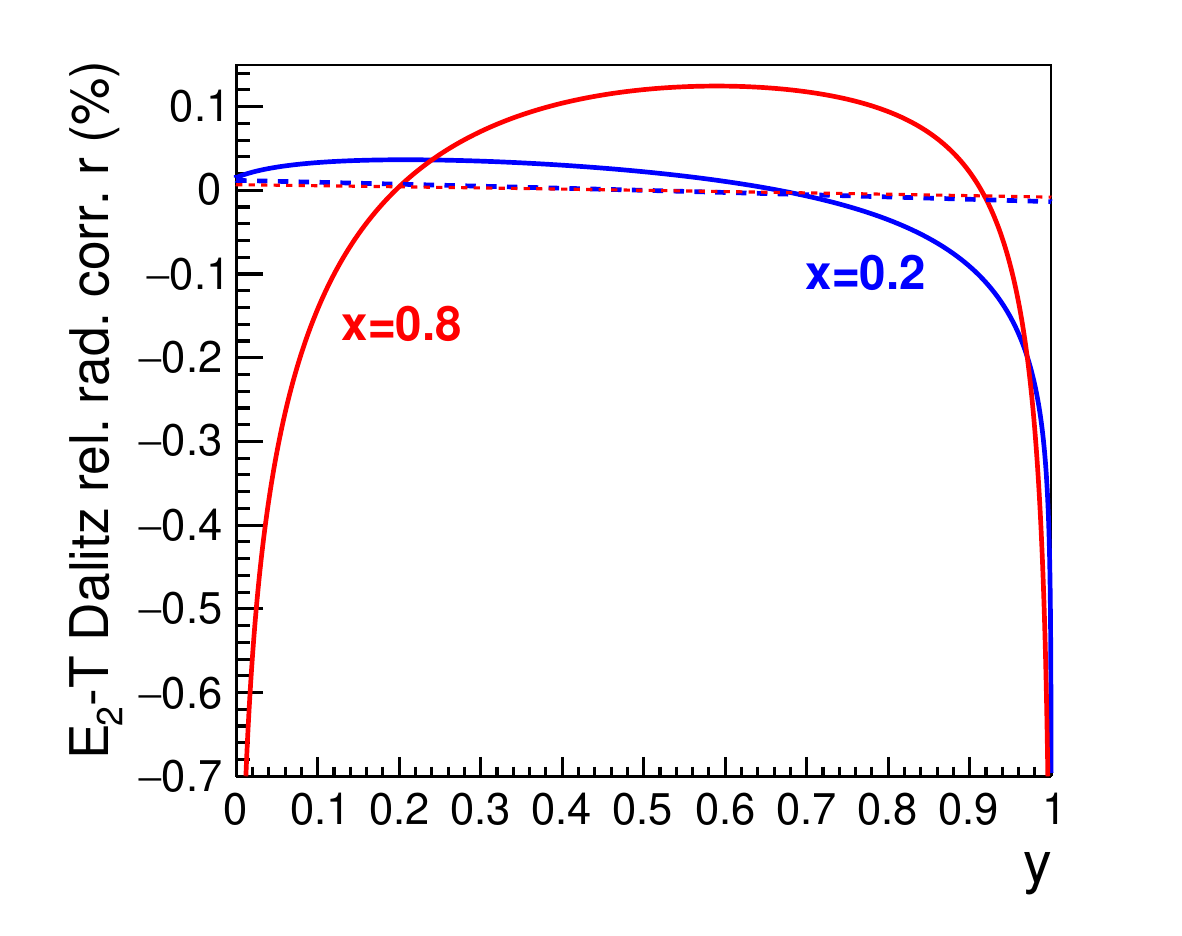}  \label{FigDalitzA}}\quad
    \subfigure[10 MeV nuclear decay]{\includegraphics[width=0.47\textwidth]
                  {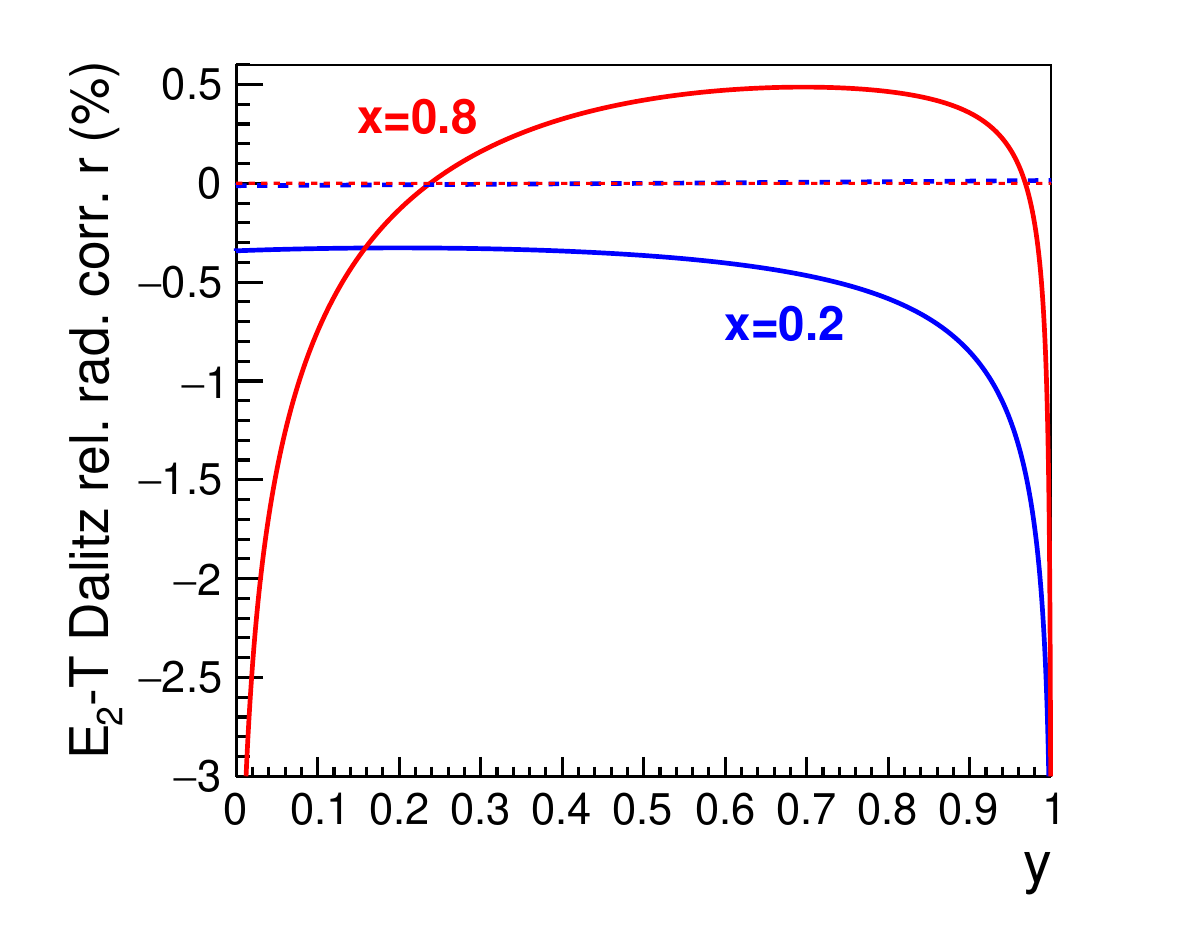}    \label{FigDalitzNB}}         
    \caption{Relative radiative correction $r$ to the $(E_2,T)$ Dalitz distribution of neutron decay (left), 
           and nuclear $\beta^-$ decay with $\Delta=10$ MeV and $m_f=20\; m_p$ (right) for
      2 different electron energies (blue: $x=0.2$, red: $x=0.8$; see Eqs. \ref{xy} and \ref{Eqr}; the blue and red dashed
     curves are almost identical). 
      Recoil-type correction: solid curves;
      neutrino-type correction: dashed curves.}
    \label{FigDalitz}
\end{figure}

The two plots in Fig.  \ref{FigDalitz} show clearly that the recoil-type and the neutrino-type
radiative corrections of the Dalitz distribution $(E_2,T)$ are completely different;
the difference increases with the decay parameter $\Delta$
(at the right-hand side, the recoil-type correction is larger while the neutrino-type one is smaller than
at the left-hand side).
There is another difference that is not visible in Fig.  \ref{FigDalitz}: 
the recoil-type bremsstrahlung correction is non-zero (positive) in the region OUT, where the neutrino-type correction
is zero (here the zeroth-order Dalitz distribution is not defined, therefore we can define only the absolute radiative correction).
It is also conspicouos that the recoil-type correction $r$ has logarithmic singularity near the lower and upper
Dalitz-plot boundaries $T=T_{min}(E_2)$ ($y=0$) and $T=T_{max}(E_2)$ ($y=1$) at $x=0.8$
(at $x=0.2$ only at the upper boundary $y=1$). 
As in the case of the electron spectrum,
these singularities are connected with the 
disappearance of the bremsstrahlung photon phase space near these boundaries.
For $E_2>E_{2h}$ (see Eq. \ref{ERpoints} in appendix \ref{SectionDalitzPlotFormulas}) at the lower
and upper boundaries, and for $E_2<E_{2h}$ only at the upper boundary, $Q_1=Q_0$  (see Eq. \ref{Qlimits}   ),
and the singularities are caused by $\ln(Q_0-Q_1)$ terms in 
the bremsstrahlung part of the radiative correction \cite{TothSzegoMargaritis1986,Gluck1986-89,GluckToth1990}.
The radiative correction has similar logarithmic singularity
in the OUT region near the E-R boundary where $Q_2=Q_0$
(note that here the virtual correction is absent); this is caused by $\ln(Q_0-Q_2)$ terms in the bremsstrahlung integrals.
These logarithmic singularities are missing in the neutrino-type radiative corrections, as it is obvious in 
Fig. \ref{FigDalitz}. Similar to the electron spectrum case, the logarithmic singularities of the Dalitz plot correction
could be eliminated by exponentiation, but this is for the experimental analyses not important.

\begin{figure}[!htbp]
    \centering
    \includegraphics[width=0.75\textwidth]{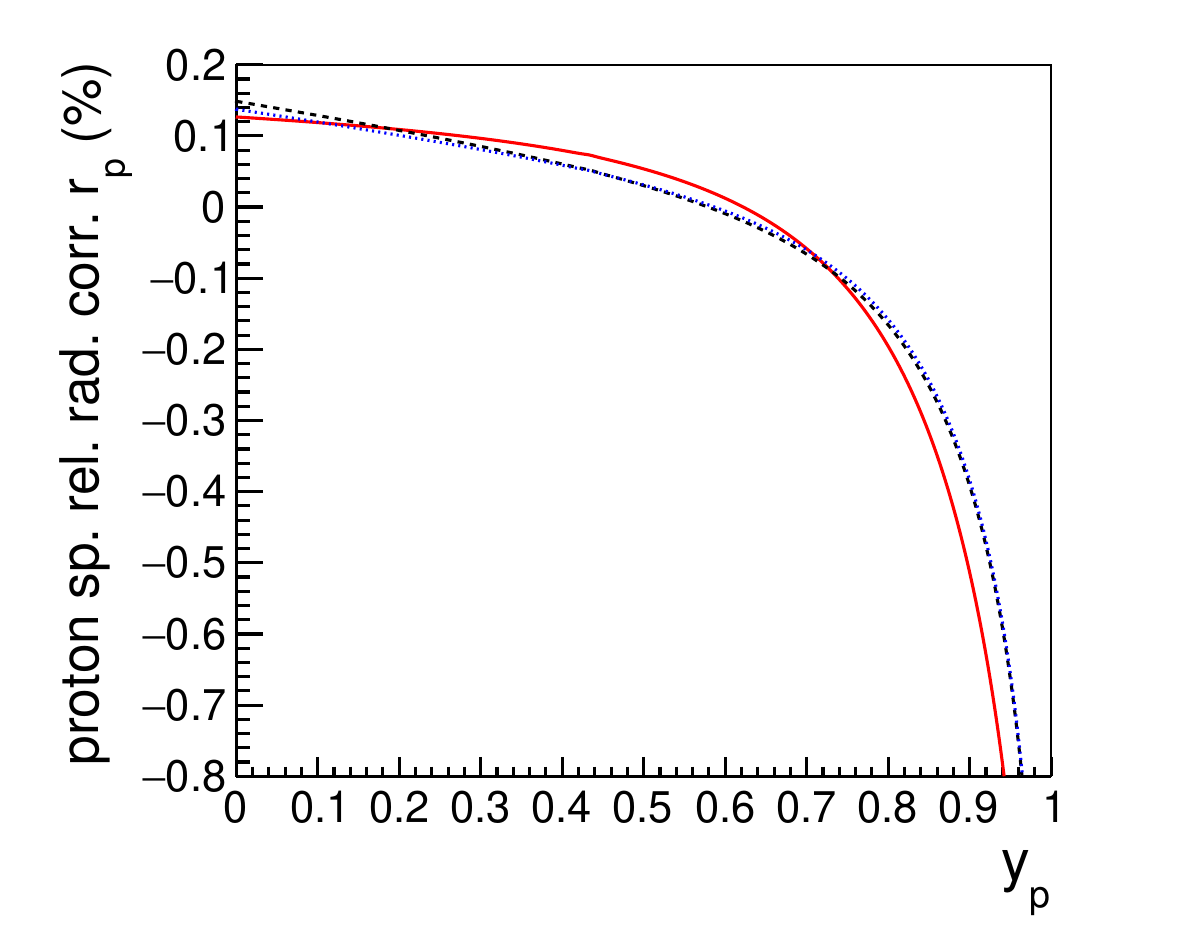}
    \caption{Relative radiative correction $r_p$ to the proton energy spectrum of neutron decay;
      $T=y_p\, T_m$.
     Recoil-type correction: solid (red) curve; neutrino-type correction: dashed (black) curve;
     hypothetical electron-type correction: dotted (blue) curve.}
    \label{FigProton}
\end{figure}

Fig. \ref{FigProton} presents the relative radiative correction $r_p$ to the proton energy spectrum 
in neutron decay, defined by
\begin{equation} \label{Eqrp}
r_p=100\frac{w_{p\gamma}(T)}{w_{p0}(T)}-r_\rho,
\end{equation}
where the zeroth-order and radiative correction proton spectra $w_{p0}(T)$ and $w_{p\gamma}(T)$
 are the integrals of $W_0(E_2,T)$ and $W_\gamma(E_2,T)$ with respect to $E_2$ 
(over the regions IN and IN+OUT, respectively),
and $r_\rho=1.503$ \% denotes the relative order-$\alpha$ MI radiative correction to the total decay rate of neutron decay:
\begin{equation} \label{Eqrrho}
r_\rho=100\frac{\int dT w_{p\gamma}(T)}{\int dT w_{p0}(T)}.
\end{equation}
Similarly to Eq. \ref{Eqr}, the total decay rate correction $r_\rho$ is subtracted in order to get smaller
$r_p$ values; we are interested here only in the proton spectrum shape.
The proton kinetic energy $T$ in Fig. \ref{FigProton} is defined by the 
horizontal axis parameter $y_p$ introduced in Eq. \ref{xy}.

In addition to the recoil-type and neutrino-type radiative corrections of the 
proton spectrum, Fig. \ref{FigProton}
contains also a hypothetical, so-called electron-type correction: this is defined by the electron
spectrum radiative correction function $r_e(x)$ of Eq. \ref{Eqre}, by assuming
that the Dalitz relative radiative correction function $W_\gamma(E_2,T)/W_0(E_2,T)$
has no $T$-dependence, i.e. using
the hypothetical  assumption that
$W_\gamma(E_2,T)=0.01\, r_e(x)\,  W_0(E_2,T)$, implying $r=0$.
One can see in  Fig. \ref{FigProton} that all 3 corrections have logarithmic singularity near 
the proton energy maximum $T=T_m$, $y_p=1$, and the neutrino-type correction function
(dashed, black curve) is rather close to the electron-type correction function (dotted, blue curve).
This is a consequence of the very small neutrino-type Dalitz correction $r$ of Eq. \ref{Eqr}
(as one can see in Fig. \ref{FigDalitz}). 
In this case, the  logarithmic singularity of the $r_e(x)$ function near $x=1$ (electron energy maximum)
causes the singularity of the proton spectrum radiative correction singularity at $y_p=1$.
The 3 correction functions have similar behaviour since a large part of them is dominated by the
similarity of the $W_\gamma(E_2,T)$ Dalitz distribution corrections and by the integration over the electron energy.

\begin{figure}[!htbp]
    \centering
    \includegraphics[width=0.75\textwidth]{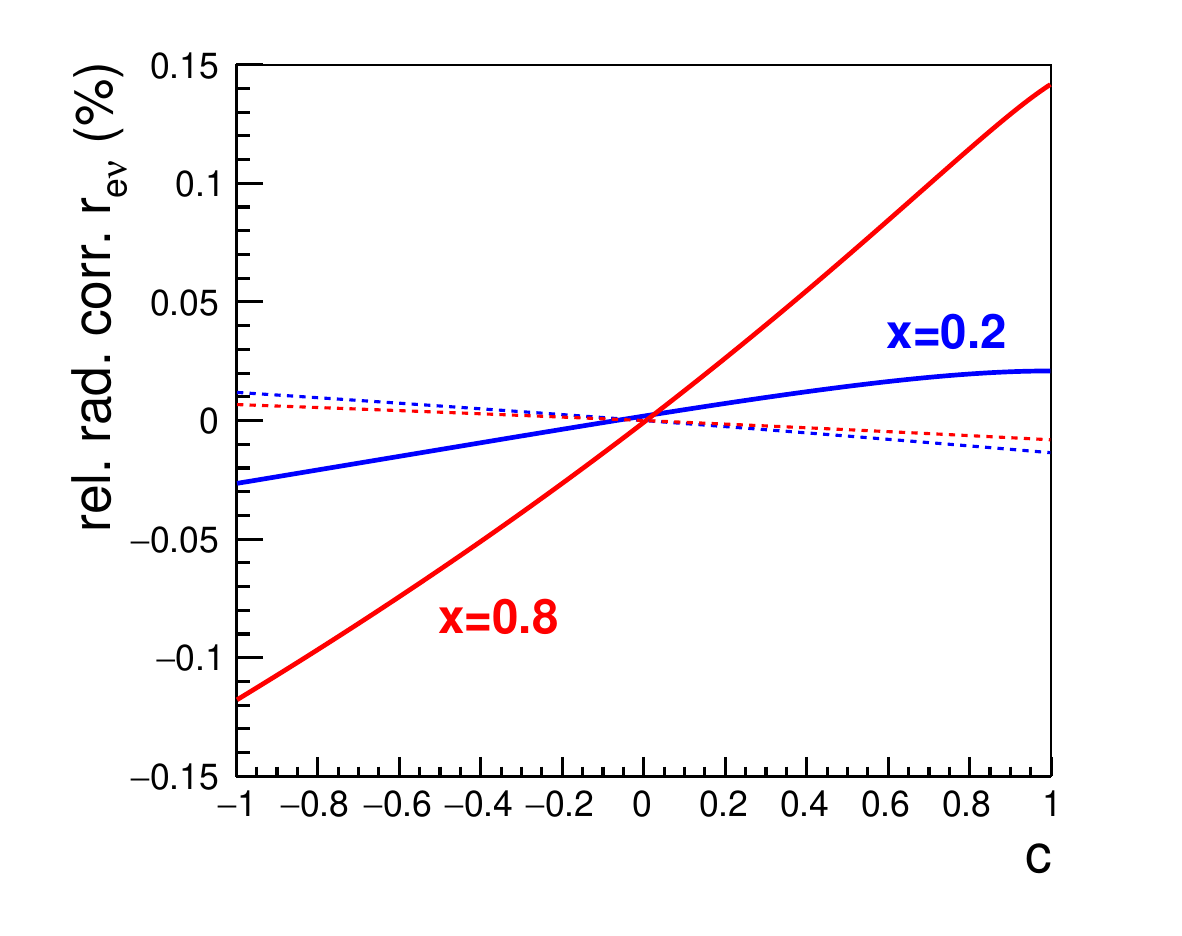}
    \caption{Relative radiative correction $r_{e\nu}$ to the $(E_2,c)$ electron-neutrino correlation
    Dalitz distribution of neutron decay, for two different electron energies 
    (blue: $x=0.2$, red: $x=0.8$; see Eq. \ref{xy}).
     Recoil-type correction: solid curves; neutrino-type correction: 
     dashed curves.}
    \label{FigEnu}
\end{figure}

Fig. \ref{FigEnu} shows examples for the radiative corrections to the $(E_2,c)$ 2-dimensional Dalitz
distribution, where $c=\cos \theta_{e\nu}$ denotes the cosine of the electron-neutrino angle.
In zeroth-order case,  $\theta_{e\nu}$ is the angle between the electron and neutrino momenta
(${\bf p_2}$ and ${\bf p_1}$, respectively), and this is also the definition for the neutrino-type
radiative correction (we use $P_2=|{\bf p_2}|$):
\begin{equation} \label{Eqp2p1}
c=c_{12}=\cos \theta_{e\nu}=\frac{{\bf p_2}\cdot {\bf p_1}}{P_2 E_1} .
\end{equation}
On the other hand, for the recoil-type radiative correction calculations we use a different,
experimentally more worthwhile,  definition:
\begin{equation} \label{Eqp2pf}
c=\cos \theta_{e\nu}=-\frac{{\bf p_2}\cdot ({\bf p_2}+ {\bf p_f}) }
   {P_2 |{\bf p_2}+{\bf p_f}|},
\end{equation}
where the momentum vector $-({\bf p_2}+{\bf p_f})$ 
of the pseudo-neutrino (defined in Sec. \ref{SectionIntroduction})
is used  instead of the neutrino momentum
 ${\bf p_1}$ (see Sec. 4 in Ref. \cite{Gluck1993}). 
In both cases, the minimal and maximal values of $c$ are $-1$ and $+1$, respectively, and
the relative radiative correction $r_{e\nu}$ is defined similarly to Eq. \ref{Eqr},
with the zeroth-order and radiative correction $(E_2,c)$ Dalitz-distributions:
\begin{equation} \label{Eqrenu}
r_{e\nu}=100\frac{W_\gamma^{e\nu}(E_2,c)}{W_0^{e\nu}(E_2,c)}-r_e(x),
\end{equation}
where
\begin{equation} \label{DalitzIntegralsenu}
\rho_0=\int\limits_{m_2}^{E_{2m}} dE_2 \int\limits_{-1}^{1} dc \cdot W_0^{e\nu}(E_2,c), \quad\quad
\rho_\gamma=\int\limits_{m_2}^{E_{2m}} dE_2 \int\limits_{-1}^{1} dc \cdot W_\gamma^{e\nu}(E_2,c).
\end{equation}

In the case of the neutrino-type radiative correction calculation, the following simple formula can be derived from 
Eq. \ref{garciamaya}:
\begin{equation} \label{Eqrenuneutrinotype}
r_{e\nu}\approx 100 \frac{\alpha}{\pi} f_n(E_2) \frac{a\beta c}{1+a\beta c},
\end{equation}
where the function $f_n(E_2)$ is defined in Eq. \ref{fn}.
This is a good example of the inconsisteny of the neutrino-type radiative corrections that we discussed in Sec.
\ref{SectionIntroduction}. Namely,  the electron-neutrino angle $\theta_{e\nu}$ has to be reconstructed
from the measured electron and recoil particle quantities (there are various possibilities for this reconstruction, depending on
experimental details), but here in the radiative correction calculation one does not use anything about this reconstruction, instead
the observation of the neutrino direction is implicitly assumed.

The recoil-type radiative correction calculation, using the definition of Eq. \ref{Eqp2pf}, is much more complicated:
one has to fix the angle between the electron momentum ${\bf p}_2$ and the pseudo-neutrino momentum
$-{\bf Q}$ in Fig. \ref{FigKinematics} (in addition to the electron energy), and one has to perform the bremsstrahlung
 integration with respect to the variable $Q$ according to
this constraint \cite{GluckToth1990,Gluck1993,GluckSANDI}.
One can see in Fig. \ref{FigEnu} that, similarly to Fig. \ref{FigDalitz}, the recoil-type 
radiative correction is much larger than the neutrino-type correction (especially for large electron energy),
and they have different slope signs : the recoil-type correction increases with $c$ 
(compare with table 5 of Ref. \cite{Gluck1993}), while the 
neutrino-type correction decreases with $c$. 
The slope of the recoil-type electron-neutrino angle radiative correction is positive in all semileptonic decays
(see table 5 in Ref. \cite{Gluck1993} and tables 2 and 6 in Ref. \cite{GluckToth1990}):
in the presence of bremsstrahlung photons $K>0$, the parameter $Q$ becomes smaller than 
in zeroth-order case $K=0$ (see Figs. \ref{FigKinematics} and \ref{FigQKplots}), and therefore 
$c$ in Eq. \ref{Eqp2pf} is increased by the presence of the photon.
On the other hand, the slope of the neutrino-type correction depends on the sign of the electron-neutrino
correlation parameter $a$ (see Eq. \ref{Eqrenuneutrinotype}).
In contrast to the function $r$ of Fig. \ref{FigDalitz}, the recoil-type correction $r_{e\nu}$ has no
logarithmic singularity at the lower and upper boundaries $c=-1$ and $c=1$.

The recoil-type radiative correction results presented above are in complete agreement with 
our old results in
tables 2, 4 and 5 of Ref. \cite{Gluck1993}, although they were computed by different codes
(C++ SANDI in this section and FORTRAN codes in Ref. \cite{Gluck1993}), and in the case 
of the electron-neutrino correlation completely different calculation methods were used \cite{GluckSANDI}.
We mention that Ref. \cite{MirrorBeta} contains recoil-type radiative correction results,
calculated by our SANDI code,  for recoil energy spectra of 19
mirror nuclear beta decays.

The radiative correction results presented above are important for the precise analyses of various
neutron decay experiments: the Dalitz distribution  $(E_2,T)$ radiative correction of Fig. \ref{FigDalitz} is
useful for the Nab experiment \cite{Nab2017}, the recoil-type proton energy spectrum correction of 
Fig. \ref{FigProton} was used in the analysis of the aSPECT experiment \cite{aSPECT2020},
and the electron-neutrino correlation radiative correction of Fig. \ref{FigEnu} (or some appropriate 
Monte Carlo computation) can be used for the analysis of the aCORN experiment \cite{aCORN2021}.
It is important that all these experiments should use the recoil-type radiative corrections;
the neutrino-type radiative corrections are not suitable for the experimental analyses.

All the above examples are for unpolarized neutron decay, because here the radiative corrections are larger than in the
the case of asymmetry observables in polarized neutron decay. Nevertheless, in the latter case one can show also significant
differences between recoil-type and neutrino-type radiative corrections. For example, Eq. \ref{garciamaya} shows that
the neutrino-type relative radiative correction to the neutrino asymmetry is strictly zero, due to the same $G(E_2)$ function
at the $PB c_1$ term as the unpolarized radiative correction Sirlin function. 
On the other hand, the recoil-type relative radiative correction of the neutrino asymmetry is small but definitely non-zero.
In fact, this is not unique: depending on the experimental details of the electron and proton observations, we get
different corrections; in Ref. \cite{Gluck1998b} we call these observables as electron-proton asymmetries.
Table 5 in Ref. \cite{Gluck1998b} shows our radiative correction results for 4 different electron-proton asymmetry functions:
all relative corrections are below 0.1 \%. Another example is the small MI (outer) radiative correction to the 
proton asymmetry \cite{Gluck1996}.
Similarly small is the analytically calculated radiative correction to the electron 
asymmetry \cite{Shann1971}; in that case, the recoil-type and neutrino-type corrections are identical 
(apart from very small recoil-order terms).

At the end of this section, we comment on the neutrino-type radiative correction results of Refs. 
\cite{IvanovPitschmann2013,IvanovHollwieser2013,IvanovPitschmannarXiv2018}.
The radiative correction formulas of the electron and proton energy Dalitz distribution and proton energy spectrum
presented in these papers agree with our neutrino-type correction calculations. It is very unfortunate that
the authors of Refs. \cite{IvanovPitschmann2013,IvanovHollwieser2013,IvanovPitschmannarXiv2018}
did not compare their radiative correction results of the $(E_2,T)$ Dalitz plot with table 2 of Ref. \cite{Gluck1993},
because then one could have seen that these two results are completely different (as it is clear from Fig. \ref{FigDalitz}).
The presentation of the radiative correction of the proton energy spectrum in Fig. 4 and table 2 of
Ref. \cite{IvanovHollwieser2013}
is not suitable to compare with the radiative correction given in table 4 of 
Ref. \cite{Gluck1993}; nevertheless, our calculation of the neutrino-type radiative correction calculation to the
proton spectrum agrees with the formulas $g_p^{(1)}$, $g_p^{(2)}$, $f_p^{(1)}$, $f_p^{(2)}$
in Sec. VI.B of Ref. \cite{IvanovHollwieser2013}.
Concerning the electron-neutrino correlation case, 
we note that Eq. 47 in Ref. \cite{IvanovHollwieser2013} is wrong: the correct form (in
consistency with the definition of $r_{e\nu}$ in Sec. 4. of Ref. \cite{Gluck1993} and with Eq. \ref{Eqrenu}) 
is Eq. \ref{Eqrenuneutrinotype} above.
Then one can see that the $r_{e\nu}$ function of the neutrino-type radiative correction 
has also $c$-dependence (which was missed by the authors of Ref.  \cite{IvanovHollwieser2013}),
 and due to the $a\approx -0.1$  value in neutron decay the correction is about 10 times smaller (and with opposite sign) than
in Fig. 3 and table 1 of Ref. \cite{IvanovHollwieser2013}, in agreement with the 
dashed curves in Fig. \ref{FigEnu}.
The authors of Refs. \cite{IvanovPitschmann2013,IvanovHollwieser2013} claim
that, with a good approximation,  Eq.  \ref{garciamaya} provides similar  radiative correction 
results than those presented in Ref. \cite{Gluck1993}. Our conclusion is just the opposite:
the neutrino-type and recoil-type radiative correction results are completely different; in several cases, the 
neutrino-type corrections are much smaller than the recoil-type  corrections.

\section{Conclusions}
 \label{SectionConclusions}

The main message of our paper is that the so-called neutrino-type radiative correction calculations to 
observables of neutron and nuclear beta decays  with measured recoil particles are not appropriate
for the precision experimental analyses.
An important part of the radiative corrections is the bremsstrahlung correction where the bremsstrahlung photons
change significantly the beta decay kinematics from 3-body to 4-body type; therefore the employment of zeroth-order
3-body kinematics is not suitable in the bremsstrahlung part of the radiative correction calculations.
In the neutrino-type radiative correction calculations the neutrino direction is fixed (in addition to the electron energy
and direction):
one can then perform analytically the bremsstrahlung integrals, and the results are rather simple.
Since the neutrino is usually not measured, the authors of the neutrino-type radiative correction publications use
the zeroth-order 3-body decay kinematics to connect the recoil particle parameters to the neutrino direction.
The main problem is: during the 
integration with respect to the bremsstrahlung photon parameters, the recoil particle
momentum has to follow the photon momentum,  due to momentum and energy conservation
(the electron momentum and neutrino direction are fixed, and the neutrino energy is constrained by 
the electron and photon energy).
Therefore, during this bremsstrahlung integration the information about the recoil particle parameters is lost
(at least at the radiative correction precision level),
the neutrino direction cannot be determined from the measured electron and recoil particle momenta,
and so the 
neutrino-type radiative corrections should not be used
for quantitities, like electron-neutrino correlation or recoil particle energy
spectrum, where the recoil particle parameters are crucial for the experimental analyses.

The correct way of the radiative corrections calculations is by fixing the recoil particle parameters (energy and direction) during
the bremsstrahlung photon integrations; we call this recoil-type radiative corrections. 
In that case, the kinematical information about the neutrino is lost, but this is 
no problem since the neutrino is usually not measured. There are several different methods of the bremsstrahlung integrations:
i, complete multidimensional numerical integration, ii, complete analytical integration, iii, semi-analytical integration
(using both analytical and numerical integrations), iv, Monte Carlo integration. In our opinion, the Monte Carlo method is
the most advantageous among the various techniques: it is relatively simple (the computer takes over large part of the
task), it is appropriate to compute radiative corrections to many different quantities (also with complicated experimental details,
like kinematic cuts or detection efficiencies), and it can be used for event generations (few hundred million events can be
generated in a few minutes).

The 4-body decay kinematics due the bremsstrahlung photon has important consequences concerning the
experimental analyses. One example is the electron and recoil particle energy Dalitz plot: due to the bremsstrahlung photons,
there is a kinematically allowed region outside  the zeroth-order Dalitz region.

We compared in the previous section the results of the neutrino-type and recoil-type radiative correction methods
for several quantities, like electron and recoil particle energy and electron-neutrino correlation Dalitz distributions, and recoil
particle energy spectrum. One can see that the results of these two radiative correction calculation types are completely different:
in several cases, the neutrino-type corrections are much smaller than the recoil-type  corrections.

\section*{Acknowledgments}

I would like to express my special gratitude to Dr. K\'alm\'an T\'oth for his introducing me to the bremsstrahlung photon kinematics problem
discussed in this paper, and for his helping me to start my scientific carrier in KFKI, RMKI, Budapest.
I would like to thank Werner Heil, Stefan Bae\ss{}ler, Ulrich Schmidt, Simon Vanlangendonck, Nathal Severijns and
Leendert Hayen for their reading the manuscript and for many useful comments, critics and discussions.
Last, but not least, I would like to thank my beloved wife Judit and our daughter Fanni
for their love, patience and perseverance during the long preparation and writing period of this paper.

\appendix

\section{Electron spectrum and asymmetry outer radiative correction formulas}
 \label{SectionRadCorrFormulas}

The electron spectrum shape outer radiative correction function $g(E_2)$ introduced by Sirlin in Ref. \cite{Sirlin1967}
can be written as
\begin{multline} \label{gE2}
g(E_2)=3\ln \left( \frac{m_p}{m_2}\right) -\frac{ 3 }{4} +4 \left[\frac{U}{\beta}-1\right] \left\{
\delta/3-\frac{ 3 }{ 2} +\ln\left[\frac{2\delta E_2}{  m_2} \right] \right\}
-\frac{ 4 }{\beta}Li_2\left( \frac{2\beta}{1+\beta}  \right) + \\
+ \frac{ U}{ \beta} \left[ 2(1+\beta^2) + \delta^2/6  -4U\right],
\end{multline}
with the proton mass $m_p$, electron velocity $\beta=\sqrt{1-m_2^2/E_2^2}$, 
the expressions $\delta$ and $U$, and the dilogarithm (Spence) function $Li_2$ \cite{Ginsberg1975}
\begin{equation} \label{ULi2}
\delta=\frac{E_{2m}-E_2}{E_2},\quad\quad   U=\frac{1}{ 2} \ln \frac{1+\beta }{1-\beta }, \quad\quad Li_2(z)=-\int\limits_0^z dt 
\frac{\ln |1-t|}{t}.
\end{equation}

The outer radiative correction function $h(E_2)$ introduced by Shann in Ref. \cite{Shann1971} (Eq. 10), which
is needed for the outer correction of the electron asymmetry, can be expressed by the function $f_n(E_2)$ of
Eq. 46 in Ref. \cite{IvanovHollwieser2013} (Eq. D-58 in \cite{IvanovPitschmannarXiv2018}):
\begin{equation} \label{fn}
h(E_2)=g(E_2)+2 f_n(E_2), \quad
f_n(E_2)=\frac{2}{3}\delta \left(1+\frac{\delta}{8}\right) \frac{1-\beta^2}{\beta^2} 
\left(\frac{U}{\beta}-1\right) -\frac{\delta^2}{12} +(1-\beta^2)\frac{U}{\beta}.
\end{equation}
The function $g_n$ of Eq. 46 in Ref. \cite{IvanovHollwieser2013} (Eq. A7 in \cite{IvanovPitschmann2013}) without the last term $C_{WZ}$
is $g_n=g(E_2)/2$.

\section{3-body decay Dalitz-plot formulas}
 \label{SectionDalitzPlotFormulas}

\begin{figure}[!htbp]
    \centering
    \includegraphics[width=0.75\textwidth]{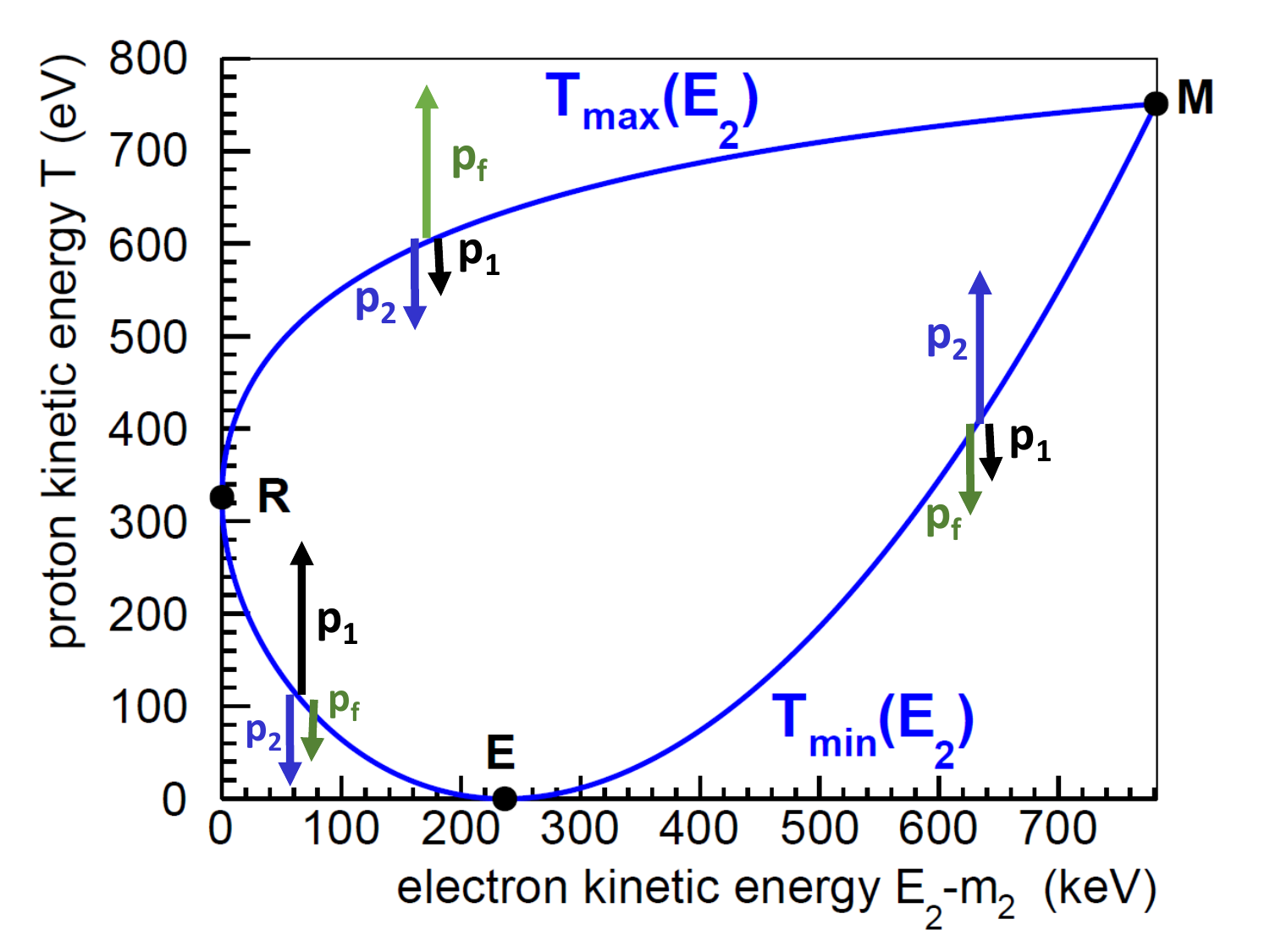}
    \caption{$(E_2-m_2,T)$ 3-body decay (zeroth-order) Dalitz plot for neutron decay, illustrated with electron (${\bf p_2}$),
        proton (${\bf p_f}$) and neutrino (${\bf p_1}$) momentum vector directions on the boundaries.}
    \label{FigDalitzPlot3}
\end{figure}

The 3-body decay Dalitz-plot upper and lower boundary curves of Fig. \ref{FigDalitzPlots} (maximal and minimal recoil 
kinetic energy $T$ as function of the electron energy $E_2$)
can be expressed as
\begin{equation} \label{Tmaxmin}
T_{max/min}(E_2)=\frac{(\Delta-E_2\pm P_2)^2}{2(m_i-E_2\pm P_2)},
\end{equation}
where $P_2=\sqrt{E_2^2-m_2^2}$ is the electron momentum (with the $c_{\rm light}=1$ notation), 
$\Delta=m_i-m_f$ is the mass difference of the decaying (initial) and daughter (final) nuclei
(in neutron decay $m_i$ and $m_f$  are the neutron and proton masses, respectively), 
and $m_2$ is the electron mass (compare with Eq.  3.6 of Ref. \cite{Gluck1993}).
With 3-body zeroth-order kinematics,  the
electron, recoil particle and neutrino direction vectors lie on a straight line on 
the upper and lower boundary curves of the Dalitz-plot; see Fig. \ref{FigDalitzPlot3}.
On the upper curve R-M of Fig.  \ref{FigDalitzPlot3} (and Fig. \ref{FigDalitzPlots}), defined 
by $T=T_{max}(E_2)$, 
 the ${\bf p}_1$ and  ${\bf p}_2$ momenta are parallel, and antiparallel to the
${\bf p}_f$ momentum vector; therefore
$P_f=P_2+P_1$, with $P_f=\sqrt{T^2+2m_f T}$ and $P_1=E_1=\Delta-E_2-T$.
On the  left-hand side lower curve R-E ($T=T_{min}(E_2)$, $E_2<E_{2h}$):
${\bf p}_f$ and ${\bf p}_2$ are parallel, and antiparallel
to ${\bf p}_1$, so $P_1=P_2+P_f$.
 On the right-hand side lower curve E-M  ($T=T_{min}(E_2)$, $E_2>E_{2h}$):
${\bf p}_f$ and ${\bf p}_1$ are parallel, and antiparallel
to ${\bf p}_2$, so $P_2=P_1+P_f$.
The Dalitz-plot points E, R and M have the following $(E_2-m_2,T)$ electron and recoil kinetic 
energy coordinates: ${\rm E}\to (E_{2h}-m_2, 0)$,  ${\rm R}\to (0, T_{h})$ and
${\rm M}\to (E_{2m}-m_2, T_{m})$, with
\begin{equation} \label{ERpoints}
E_{2h}=0.5\left(\Delta +m_2^2/\Delta\right), \quad T_{h}=(\Delta-m_2)^2/(2(m_i-m_2)),
\end{equation}
\begin{equation} \label{Mpoint}
E_{2m}=\Delta-T_m, \quad T_{m}=(\Delta^2-m_2^2)/(2m_i)
\end{equation}
(compare with Eqs. 3.8, 3.10 and 3.3 of Ref. \cite{Gluck1993}, and with Eqs. A5-A6 of Ref. \cite{Gluck1998a}).

The E-M and E-R-M Dalitz-plot curves  of Figs.  \ref{FigDalitzPlots} and \ref{FigDalitzPlot3}
can also be expressed by the functions $E_{2max}(T)$
and $E_{2min}(T)$:
\begin{equation} \label{E2maxmin}
E_{2max/min}(T)=\frac{(\Delta-T\pm P_f)^2+m_2^2}{2(\Delta- T\pm P_f)}
\end{equation}
(compare with Eq.  3.7 of Ref. \cite{Gluck1993}).

\section{Bremsstrahlung integral derivations}
 \label{SectionDerivation}

The bremsstrahlung total decay rate can be expressed with the general 4-body decay 
phase space integral formula (see Eqs. 2.5-2.6 in Sec. III. of Ref. \cite{BycklingKajantie}):
\begin{equation} \label{phasespaceintegral}
\rho_{BR}=C\int \frac{d^3 {\bf p}_2}{E_2}  \frac{d^3 {\bf p}_1}{E_1}  \frac{d^3 {\bf p}_f}{E_f}
 \frac{d^3 {\bf k}}{K_0} \delta^4(p_i-p_1-p_2-p_f-k)\cdot M_{BR},
\end{equation}
with
\begin{equation} \label{MBR}
M_{BR}=\sum_{2,f,k} |{\cal M}_{BR}|^2, \quad C=\frac{1}{2^{13} \pi^8 m_i}, \quad
K_0=\sqrt{K^2+m_\gamma^2},
\end{equation}
the index $i$ denotes the decaying (initial) particle, and 
the bremsstrahlung amplitude squared is summed over the polarization states of the final particles.

\subsection{Recoil-type calculation}
 \label{SectionDerivationRecoilType}

In the case of the recoil-type radiative correction calculation, the goal is to keep the fixed recoil particle 
momentum, and to get rid of the unobserved neutrino momentum. Therefore, one uses
the Dirac-delta identity (see Eq. III.2.11 in Ref. \cite{BycklingKajantie})
\begin{equation} \label{DiracDeltaNeutrino}
\frac{d^3 {\bf p}_1}{E_1} = 2 \Theta(E_1) \delta(p_1^2) d^4 p_1,
\end{equation}
together with (using $PdP=EdE$)
\begin{equation} \label{d3p}
\frac{d^3 {\bf p}_2}{E_2}=P_2 dE_2 d\Omega_2, \quad 
\frac{d^3 {\bf p}_f}{E_f}=P_f dE_f d\Omega_f,\quad
\frac{d^3 {\bf k}}{K_0}=K dK_0 d\Omega_k.
\end{equation}
We get  the following integral containing still 1 Dirac-delta:
\begin{equation} \label{integralnext}
\rho_{BR}=2C\int P_2dE_2 d\Omega_2 \cdot P_f dE_f d\Omega_f \cdot KdK_0  d\Omega_k \cdot
\Theta(E_1) \delta (p_1^2)\cdot M_{BR}.
\end{equation}
Here the neutrino 4-momentum and energy are expressed by the other particle 4-momenta and energies:
$p_1=p_i-p_2-p_f-k$, $E_1=m_i-E_2-E_f-K_0$,
and the differential solid angles can be expressed by polar and azimuthal angles, like
$d\Omega_k=d\cos\theta_k d \phi_k$.
The remaining Dirac-delta can be eliminated by integration over the photon polar
angle $\theta_k$:
\begin{equation} \label{dcosthetak}
p_1^2=Q_0(Q_0-2K)-Q^2-2QK\cos\theta_k, \quad \int d\cos\theta_k  \delta (p_1^2) 
\Rightarrow 1/(2QK).
\end{equation}
The $\cos\theta_k=\pm 1$ limits define here the photon momentum magnitude limits in Eq. \ref{Klimits}.

The polar angle $\theta_f$ of the recoil particle can be defined relative to the electron momentum
${\bf p}_2$, and by the cosine theorem we get:
$Q^2=P_2^2+P_f^2+2P_2P_f\cos\theta_f$. Therefore, the integration by $d\Omega_f$ can be replaced by
the integration over $Q$ and $\phi_f$:
\begin{equation} \label{dOmegaf}
d\Omega_f \Rightarrow \frac{Q}{P_2 P_f} dQ\, d\phi_f.
\end{equation}
With the $\cos\theta_f=\pm 1$ limits in the above cosine theorem equation
we get the $Q_1\le Q\le Q_2$ limits of Eq. \ref{Qlimits} in region OUT (se  Fig. \ref{FigQKplotB}).
In addition, the invariant mass squared of the pseudo-neutrino according to Eq. \ref{ConservationEquations}
should be non-negative:
\begin{equation} \label{PseudoNeutrino}
(p_1+k)^2=Q_0^2-Q^2=2(p_1 k)=2E_1 K (1-\cos \theta_{1k})\ge 0,
\end{equation}
therefore we get the upper $Q$-limit in region IN (where $Q_2>Q_0$): $Q\le Q_0$  (see Fig. \ref{FigQKplotA}).

We get then the integral
\begin{equation} \label{lastintegral}
\rho_{BR}=C\int dE_2 d\Omega_2 dE_f d\phi_f dK d\phi_k dQ \Theta(E_1)\cdot   K/K_0 \cdot M_{BR}.
\end{equation}
In the unpolarized case: $\int d\Omega_2 d\phi_f \to 8\pi^2$, and so we get
Eq. \ref{WBR}.

In the above derivation we used zero mass values for both the photon and the neutrino. Nevertheless, for the infrared 
regularization one should use finite photon mass $m_\gamma$, and in that case the photon momentum limits are different from Eq. 
\ref{Klimits}. The neutrino 4-momentum squared is then
\begin{equation} \label{p1squared}
p_1^2=Q_0^2-Q^2+m_\gamma^2-2Q_0 K_0-2Q K \cos\theta_k,
\end{equation}
and from  the $\cos\theta_k=\pm 1$ limits we get, with $\delta=Q_0-Q$, $\; S=Q_0+Q$:
\begin{equation} \label{KminKmax}
K_{max}=\frac{S^2-m_\gamma^2}{2S}, \quad
K_{min}=
\begin{cases} 
 (\delta^2-m_\gamma^2)/(2\delta) &: \quad \text{for} \quad \delta>m_\gamma, \\
 (m_\gamma^2-\delta^2)/(2\delta) &: \quad \text{for} \quad \delta<m_\gamma. 
 \end{cases}
\end{equation}
In addition, the maximum value of $Q$ in the region IN
is now $Q_{max}=\sqrt{Q_0^2-m_\gamma^2}$, instead of $Q_0$.

\subsection{Neutrino-type calculation}
 \label{SectionDerivationNeutrinoType}

In the case of the neutrino-type radiative correction calculation, the goal is to keep the neutrino
direction as fixed quantity, and to get rid of the recoil particle momentum. Therefore, one uses
the following Dirac-delta identity:
\begin{equation} \label{DiracDeltaRecoil}
\frac{d^3 {\bf p}_f}{E_f} = 2 \Theta(E_f) \delta(p_f^2-m_f^2) d^4 p_f.
\end{equation}
In the following, we neglect the recoil-order terms in the bremsstrahlung integrals (in Ref. \cite{GluckGENDER}
we present the kinematically exact formulas, where the recoil-order terms are not neglected).
Using $p_f^2\approx m_i^2-2 m_i(E_1+E_2+K)$ and $m_i^2-m_f^2\approx 2m_i \Delta$:
\begin{equation} \label{DiracDeltaRecoilApprox}
\frac{d^3 {\bf p}_f}{E_f} \approx  \frac{1}{m_i} \delta(\Delta-E_1-E_2-K) d^4 p_f.
\end{equation}
Using Eqs. \ref{d3p}, we get
\begin{equation} \label{integralneutrinotype}
\rho_{BR}\approx \frac{C}{m_i}  \int\limits_{m_2}^\Delta dE_2 
\int\limits_{4\pi} d\Omega_2 \int\limits_{4\pi} d\Omega_1  \int\limits_{4\pi} d\Omega_k
\int_0^{\Delta-E_2} dK  \frac{K^2}{K_0} P_2 (\Delta-E_2-K) \cdot M_{BR}
\end{equation}
(compare with Eq. 4.10 of Ref. \cite{Gluck1997}). 
This multidimensional integral has very simple integration limits, which makes the analytical integration
much easier than in the case of the recoil-type calculation.

\end{document}